\documentclass[superscriptaddress,aps,prb,reprint]{revtex4-1}

\newcommand{\bra}[1]{\langle #1 |}			
\newcommand{\ket}[1]{| #1 \rangle}

 \newcommand{\sgn}{\operatorname{sgn}}

\renewcommand{\Re}{\operatorname{Re}}
\renewcommand{\Im}{\operatorname{Im}}

\usepackage{commath}

\usepackage[table]{xcolor}
\usepackage{amsmath,amssymb}
\usepackage[utf8]{inputenc}
\usepackage{graphicx}
\usepackage[caption=false,subrefformat=parens,labelformat=parens]{subfig}
\usepackage{epstopdf}
\usepackage{siunitx}
\usepackage{tikz}
\usepackage{placeins}
\usepackage{cleveref}

\begin{document}
\begin{abstract}
Based on group theoretical arguments we derive the most general Hamiltonian for the $\text{Bi}_2\text{Se}_3$-class of materials including terms to third order in the wave vector, first order in electric and magnetic fields, first order in strain and first order in both strain and wave vector. We  determine analytically the effects of strain on the electronic structure of $\text{Bi}_2\text{Se}_3$. For the most experimentally relevant surface termination we analytically derive the surface state spectrum, revealing an anisotropic Dirac cone with elliptical constant energy counturs giving rise to different velocities in different in-plane directions. The spin-momentum locking of strained $\text{Bi}_2\text{Se}_3$ is shown to be modified and for some strain configurations we see a non-zero spin component perpendicular to the surface. Hence, strain control can be used to manipulate the spin degree of freedom via the spin-orbit coupling. We show that for a thin film of  $\text{Bi}_2\text{Se}_3$ the surface state band gap induced by coupling between the opposite surfaces changes opposite to the bulk band gap under strain. Tuning the surface state band gap by strain, gives new possibilities for the experimental investigation of the thickness dependent gap and optimization of optical properties relevant for, e.g., photodetector and energy harvesting applications. We finally derive analytical expressions for the effective mass tensor of the Bi$_2$Se$_3$ class of materials as a function of strain and electric field.
\end{abstract}

\title{Symmetry analysis of strain, electric and magnetic fields in the $\text{Bi}_2\text{Se}_3$-class of topological insulators}

\author{Mathias Rosdahl Jensen}
\affiliation{Technical University of Denmark, Department of Photonics Engineering, Kgs-Lyngby, 2800, Denmark}

\author{Jens Paaske}
\affiliation{Center for Quantum Devices, Niels Bohr Institute, University of Copenhagen,
Universitetsparken 5, DK-2100 Copenhagen, Denmark}

\author{Anders Mathias Lunde}
\affiliation{Center for Quantum Devices, Niels Bohr Institute, University of Copenhagen,
Universitetsparken 5, DK-2100 Copenhagen, Denmark}

\author{Morten Willatzen}
\affiliation{Beijing Institute of Nanoenergy and Nanosystems, Chinese Academy of Sciences, Beijing, China \\ 
and Technical University of Denmark, Department of Photonics Engineering, Kgs-Lyngby, 2800, Denmark}

\maketitle

\section{Introduction}
Topological insulators have an inverted band gap which engenders topologically protected surface states (SS). Exhibiting linear dispersion, the electrons at the surface resemble massless helical Dirac fermions, with spin locked to the momentum. The prime examples of three-dimensional topological insulators are among $\text{Bi}_2\text{Se}_3$-class of materials \cite{oldmodel}.  This class, also known as the tetradymite group, contains compounds $\text{M}_2\text{X}_3$ where M is either Bi or Sb and X is a combination of Se, S and Te. The crystal structure consists of unit layers of five atomic layers, so-called quintuple layers. For the simplest case of a $(111)$ surface termination, where the surface is parallel to the quintuple layer, the Dirac cone is fully isotropic for small in-plane momentum, perturbed only by a hexagonal warping to third order in momentum\cite{warping}. For other surface terminations the Dirac cone of the surface states becomes anisotropic with elliptical curves of constant energy, as reported for the $(221)$ surface in Ref.~\onlinecite{sidesurfacestate}. Anisotropic Dirac fermions have interesting transport properties due to the different group velocity in different directions \cite{anisotropic_dirac_transport}, and have attracted attention in other Dirac materials \cite{Feng_diraccones,ani_dirac_graphene}.

For both fundamental research and applications it is interesting to be able to tune the physical properties of these materials. One way to do this is by the application of strain. It was reported in Refs.~\onlinecite{luo,young} that strain strongly affects the band gap at the $\Gamma$ point of bulk $\text{Bi}_2\text{Se}_3$, and could even close the gap at large strain values, i.e. induce a topological phase transition. Whether this is possible to achieve in real materials is not known, but a recent study points to the feasibility of strain induced effects in $\text{Bi}_2\text{Se}_3$ where strain up to $3\%$ was induced by lattice mismatch \cite{strain_experiment}.

In this paper, we use the method of invariants~\cite{willatzen} to derive the most general Hamiltonian at the $\Gamma$ point to third order in wave vector, first order in strain, first order in electric and magnetic fields as well as terms to first order in both wave vector and strain. The allowed third order terms in the wave vector include three terms that were neglected in a previous analysis \cite{model}. Since this model is based solely on the symmetry of the crystal, it is valid for all materials in the $\text{Bi}_2\text{Se}_3$ class.

\begin{figure}
{\includegraphics[width=\columnwidth]{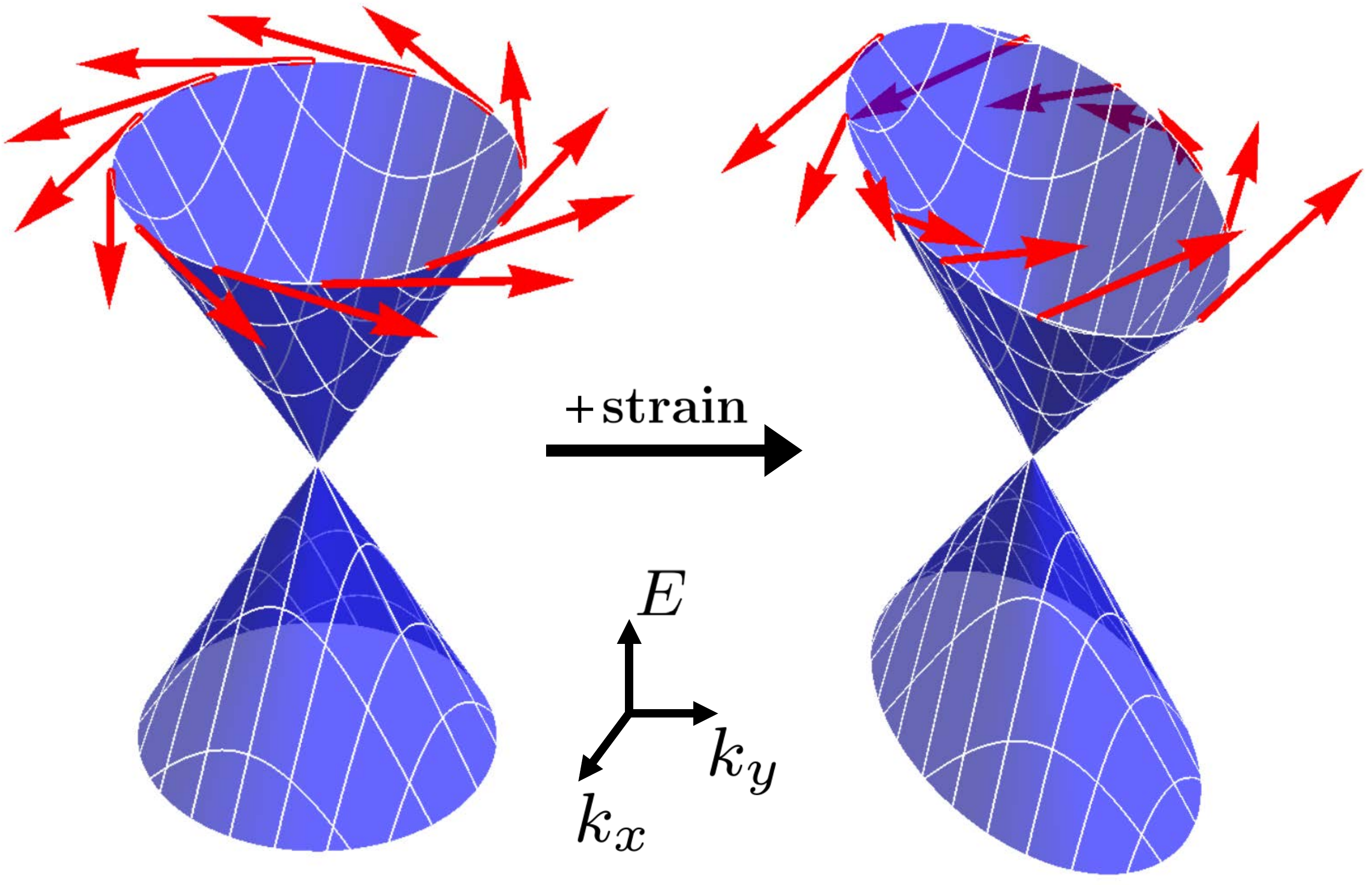}}
\caption{For unstrained $\text{Bi}_2\text{Se}_3$ the linear dispersion of the helical surface states is isotropic in the plane to linear order in the wave vector. In the case of strain, the in plane rotational symmetry is broken by terms to first order in both strain and wave vector leading to a Dirac cone with elliptical contours of constant energy, here shown for $\epsilon_{xy}=5\%$. Hence, the group velocity of the surface electrons becomes dependent on the direction. The broken rotational symmetry also allows for a spin component perpendicular to the surface, as shown by the red arrows.}
\label{fig:DiracCone}
\end{figure}

Using this model, we determine analytical expressions for the modified bulk band structure and the effective mass tensor near the $\Gamma$-point. Changes in the effective mass tensor with strain and electric field are important for transport and optical properties~\cite{jensen}. Comparing the strain dependence of the band gap to recent density functional theory (DFT) calculations, allows us to determine some of the strain related parameters in the band structure. From this bulk spectrum, we go on to investigate the effects of strain on the surface states of a semi-infinite topological insulator. We analytically derive the spectrum of the surface states, applying hard-wall boundary conditions at the surface for a semi-infinite topological insulator with a (111) surface termination. In contrast to the unstrained case, the full in-plane rotational symmetry to linear order in wave vector is broken by terms linear in both strain and wave vector, leading to an anisotropic Dirac spectrum. The spin expectation values of the SS are calculated, and it is shown that the spin-momentum locking is affected by strain, revealing a non-zero spin component perpendicular to the surface. The modified Dirac cone and spin structure of the SS is shown in Fig.~\ref{fig:DiracCone}. Finally, we show that strain affects the localization of the SS wave functions and thereby the band gap arising in thin films due to the coupling between SS on opposite surfaces. We show that the SS band gap changes oppositely to the bulk band gap under strain: increasing the bulk band gap, localizes the SS and decreases the SS band gap, and vice versa. This shows that not only the bulk, but also the SS band gap can be tuned via strain.

\section{\label{model}Model Hamiltonian}
In line with Ref.~\onlinecite{model}, we wish to derive a 4-band model Hamiltonian describing the $\text{Bi}_2\text{Se}_3$ class, which now includes all allowed terms to orders cubic in the wave vector $\mathbf k$, linear in the strain tensor $\epsilon_{ij}$, linear in magnetic field $\mathbf B$ or electric field $\mathbf E$, and linear in both $\mathbf k$ and $\epsilon_{ij}$.

In general any $4\times 4$ Hamiltonian can be written in terms of Dirac $\Gamma$ matrices defined in terms of the Pauli matrices by:
\begin{gather}
\Gamma_1=\sigma_1\otimes \tau_1, \quad \Gamma_2=\sigma_2\otimes\tau_1, \quad \Gamma_3=\sigma_3\otimes\tau_1,\nonumber\\
\Gamma_4=\sigma_0 \otimes \tau_2, \quad \Gamma_5 =\sigma_0 \otimes\tau_3,
\end{gather}
and their commutators $\Gamma_{ij} = \tfrac{1}{2i} [\Gamma_i,\Gamma_j]$. A general $4\times4$ matrix can then be written:
\begin{align}
H_m = \varepsilon \mathbf I + \sum_{i=1}^5 d_i \Gamma_i + \sum_{i,j} d_{ij} \Gamma_{ij},
\end{align}
where the coefficients  $\varepsilon,d_i,d_{ij}$ must be real to ensure hermiticity. In the present case the coefficients are functions of $\mathbf k$, $\epsilon_{ij}$, $\mathbf B$ and $\mathbf E$.

Here we use the basis $\ket{P1_-^+,\tfrac{1}{2}}$, $-i\ket{P2_+^-,\tfrac{1}{2}}$, $\ket{P1_-^+,-\tfrac{1}{2}}$, $i\ket{P2_+^-,-\tfrac{1}{2}}$, derived from the $p$ orbitals of the Bi and Se atoms. The upper sign denotes the inversion eigenvalue and the $\pm \tfrac{1}{2}$ the total angular momentum. For a more complete discussion see Ref.~\onlinecite{model}. In this basis the $\sigma$ matrices do not represent spin, but are related to spin by:
\begin{align}
s_x = \frac{\sigma_1\otimes\tau_3}{2} \quad s_y = \frac{\sigma_2\otimes\tau_3}{2} \quad s_z = \frac{\sigma_3\otimes\tau_0}{2},
\end{align}
as noted in Ref.~\onlinecite{Silvestrov}.

In this basis, the transformation operators corresponding to the symmetries of the crystal are inversion $I=\sigma_0 \otimes \tau_3$, three-fold rotation around the $z$-axis
$R_3 = e^{i\pi \sigma_3 / 3} \otimes \tau_0$,
two fold rotation around the $x$-axis $R_2 = i \sigma_1 \otimes \tau_3$ and time-reversal $T= i(\sigma_2 \otimes \tau_0)K$, where $K$ is the complex conjugation operator. Now the $\Gamma$-matrices can be characterized by their irreducible representation under the group $D_3$ as well as their eigenvalues under $I $ and $T$.

\subsection{Group theory}
Next, we construct polynomials in wave vector, strain, electric and magnetic fields transforming according to the irreducible representations of the group $D_{3d}$. This group is a direct product group of $D_3$ with the group of the inversion operator. Here we will consider $D_3$ separately and simply add inversion eigenvalues as well as time-reversal eigenvalues. The group $D_3$ has three irreducible representations: $\Gamma^{(1)}$, which is the trivial representation, $\Gamma^{(2)}$, which is one-dimensional as well and the two-dimensional representation $\Gamma^{(3)}$.

\begin{table}
    \subfloat[\label{table:multiplicationtable}]{
      \centering
        \begin{tabular}{|c|c|c|c|}
\hline
 $\otimes$ & $\Gamma^{(1)}$ & $\Gamma^{(2)}$ & $\Gamma^{(3)}$ \\ \hline
$\Gamma^{(1)}$ & $\Gamma^{(1)}$ & $\Gamma^{(2)}$ & $\Gamma^{(3)}$\\ \hline
$\Gamma^{(2)}$ & $ \Gamma^{(2)}$ & $\Gamma^{(1)}$ & $\Gamma^{(3)}$\\ \hline
$\Gamma^{(3)}$ & $\Gamma^{(3)}$ & $\Gamma^{(3)}$ & $\Gamma^{(1)} \oplus \Gamma^{(2)} \oplus \Gamma^{(3)}$ \\\hline
\end{tabular}
}
    \subfloat[\label{table:coupling23}]{
      \centering
   \begin{tabular}{|c|c|c|}
\hline
  & $u^{(2)}v^{(3)}_1$ & $u^{(2)}v^{(3)}_2$ \\ \hline
$\psi^{(3)}_1$ & i & 0  \\
$\psi^{(3)}_2$ & 0 & -i \\  \hline
\end{tabular}
}
\\
    \subfloat[\label{table:coupling33}]{
    \begin{tabular}{|c|c|c|c|c|}
\hline
  & $u^{(3)}_1v^{(3)}_1$ & $u^{(3)}_1v^{(3)}_2$ & $u^{(3)}_2v^{(3)}_1$ & $u^{(3)}_2v^{(3)}_2$ \\ \hline
$\psi^{(1)}$ & $0$ & $\frac{1}{\sqrt{2}}$ & $\frac{1}{\sqrt{2}}$  & $0$ \\ \hline
$\psi^{(2)}$ & $0$ & $\frac{i}{\sqrt{2}}$ & $\frac{-i}{\sqrt{2}}$ & 0 \\ \hline
$\psi^{(3)}_1$ & 0 & 0 & 0 & 1 \\
$\psi^{(3)}_2$ & 1 & 0 & 0 & 0 \\  \hline
\end{tabular}
}
  \caption{Multiplication table for the irreducible representations of the group $D_{3d}$, (a),  and the coupling constants for the basis functions of the relevant products $\Gamma^{(2)}\otimes\Gamma^{(3)}$, (b) and $\Gamma^{(3)}\otimes\Gamma^{(3)}$, (c). Note that the multiplication table is completely general since it is in principle concerning equivalence classes of representations, while the coupling constants are only valid for a specific representation, here the representation with basis functions $\{k_+,k_-\}$. Adopted from Ref.~\onlinecite{kostertables}.}
  \label{table:kostertables}
\end{table}

The wave vector component $k_z$ transforms according to the one-dimensional representation $\Gamma^{(2)}$, i.e. it only changes by multiplication of a constant under the transformations of $D_{3}$. The in-plane components, $k_x$ and $k_y$, transform according to the two-dimensional irreducible representation $\Gamma^{(3)}$. Here we will use the basis $\{k_+,k_-\}$, where $k_\pm=k_x\pm i k_y$. As an example we will show how to construct the second order terms transforming according to irreducible representations including only in-plane momentum. First we note that
\begin{align}
\Gamma^{(3)} \otimes \Gamma^{(3)} = \Gamma^{(1)} \oplus \Gamma^{(2)} \oplus \Gamma^{(3)},
\end{align}
i.e. the second order terms of in-plane momenta transform according to a representation equivalent to the sum of these three irreducible representations. The basis functions for the irreducible representations can be formed using Table~\subref*{table:coupling33}, with $\{u_1,u_2\}=\{v_1,v_2\} = \{k_+,k_-\}$, and we get:
\begin{align}
\Gamma^{(1)}&: \quad \psi^{(1)} = \frac{1}{\sqrt{2}} k_+k_- + \frac{1}{\sqrt{2}} k_-k_+\propto k_{||}^2, \\
\Gamma^{(2)}&: \quad \psi^{(2)} = \frac{i}{\sqrt{2}} k_+k_- + \frac{-i}{\sqrt{2}} k_-k_+ = 0, \\
\Gamma^{(3)}&: \quad \{\psi^{(3)}_1,\psi^{(3)}_2 \} = \{k_-^2,k_+^2\},
\end{align}
where we have defined $k_{||}^2 = k_x^2 + k_y^2$. Since $\{u_1,u_2\}=\{v_1,v_2\}$ the $\Gamma^{(2)}$ term is simply 0, and we have 3 independent second order terms in the in-plane momenta: $k_{||}^2$, which is invariant, and the pair $\{k_-^2,k_+^2\}$ transforming according to the two-dimensional irreducible representation $\Gamma^{(3)}$. The inversion and time-reversal eigenvalues follow the simple rule $\pm\otimes \pm =+$ and $\pm\otimes \mp =-$. Proceeding like this we can go to any desired order, and include external fields as well. We note that the strain tensor $\epsilon_{ij}$ transforms as $k_ik_j$\cite{BirPikus}. This result follows from the definition of strain
\begin{align}
 \epsilon_{ij} & = \frac{1}{2}\left( \frac{\partial u_i}{\partial x_j} + \frac{\partial u_j}{\partial x_i}\right),
\end{align}
where ${\bf u}$ is the displacement vector.
The resulting irreducible basis functions are summarized in Table~\ref{table:transformation}.

\begin{table}
 \begin{tabular}{|c|c|c|c|c|}\hline
 $T$ & $I$ & $\Gamma^{(1)}$ & $\Gamma^{(2)}$ & $\Gamma^{(3)}$  \\\hline
  & & dim=1 & dim=1 & dim=2 \\ \cline{3-5}
 - & - &
 $\begin{matrix} \Re(k_+^3) \\ \Im(k_+\epsilon_{z-}) \\ \Re(k_+\epsilon_+) \\ \Gamma_4\end{matrix}$  &
 $ \begin{matrix}k_z \\ k_z^3 \\ k_{||}^2k_z \\ \Im(k_+^3) \\ \epsilon_{zz} k_z \\ \epsilon_{||} k_z \\  \Re(k_+\epsilon_{z-}) \\ \Im(k_+\epsilon_+) \\ \Gamma_3\end{matrix}$ &
 $\begin{matrix}\{k_+,k_-\}\\ k_{||}^2 \{k_+,k_-\} \\ k_z^2\{k_+,k_-\} \\ \{ik_z k_-^2, -ik_z k_+^2\} \\ \epsilon_{zz}\{k_+,k_-\} \\ \epsilon_{||}\{k_+,k_-\} \\ k_z\{\epsilon_{z+},\epsilon_{z-}\} \\ k_z\{i\epsilon_-,-i\epsilon_+\} \\  \{ik_-\epsilon_{z_-},-ik_+\epsilon_{z+}\} \\ \{k_-\epsilon_+,k_+\epsilon_-\} \\ \{\Gamma_+^{1,2},\Gamma_-^{1,2}\}\end{matrix}$
 \\\hline
+ & +  &
$\begin{matrix}k_z^2 \\ k_{||}^2 \\ \epsilon_{zz} \\ \epsilon_{||}  \\ \Gamma_5\end{matrix}$ &
  &
  $\begin{matrix} \{ik_z k_+, -ik_z k_-\} \\ \{k_-^2,k_+^2\} \\ \{i\epsilon_{z+},-i\epsilon_{z-}\} \\ \{\epsilon_{-},\epsilon_{+}\} \end{matrix}$ 	
  \\\hline
  + & - &
  $\begin{matrix} \Gamma_{45}\end{matrix}$ &
  $\begin{matrix} E_z \\ \Gamma_{35}\end{matrix}$ &
$ \begin{matrix} \{E_+,E_-\} \\ \{i\Gamma_+^{25,51},-i\Gamma_-^{25,51}\}\end{matrix}$
  \\\hline
    - & + &
  &
   $\begin{matrix} B_z \\ \Gamma_{12} \\ \Gamma_{34} \end{matrix}$ &
  $\begin{matrix} \{B_+,B_-\} \\ \{\Gamma_+^{13,23},\Gamma_-^{23,31}\} \\ \{\Gamma_+^{14,24},\Gamma_-^{14,24}\}\end{matrix}$
  \\\hline
 \end{tabular}
\caption{Polynomials of $\mathbf k$, strain and magnetic fields and the $\Gamma$ matrices under the transformations of the group $D_{3}$, inversion and time reversal. To simplify the notation we have introduced $\epsilon_{||} = \epsilon_{xx} +\epsilon_{yy}$, $\epsilon_{\pm} = \epsilon_{xx} - \epsilon_{yy} \pm 2i\epsilon_{xy}$, $\epsilon_{z\pm} = \epsilon_{zx} \pm i\epsilon_{zy}$. In the 2-dimensional representations we have changed basis to $\Gamma^{i,j}_\pm = \Gamma_i \pm i \Gamma_j$. These have been constructed such that they are real, and each pair is hermitian conjugates of each other and the pairs transform like  $\{k_+,k_-\}$ under the group $D_{3}$. }
\label{table:transformation}
\end{table}

\subsection{Model Hamiltonian} For the Hamiltonian to be invariant under the group $D_3$ it can only contain terms from the invariant representation, $\Gamma^{(1)}$. As seen in the multiplication Table~\subref*{table:multiplicationtable}, $\Gamma^{(1)}$ terms can only be formed by combining terms from the same irreducible representation. For invariance under inversion and time-reversal, we must combine terms and matrices with the same eigenvalues, i.e. from the same cell in Table~\ref{table:transformation}. For the $\Gamma^{(3)}$ representation we can use Table~\subref*{table:coupling33} to construct an invariant term, $\Gamma^{(1)}$ and $\Gamma^{(2)}$ are 1 dimensional so we simply take the product.

For example, the terms $\{i k_z k_-^2,-ik_z k_+^2\}$ belong to the irreducible representation $\Gamma^{(3)}$ and are odd under both TR and I. The pair of matrices $\{\Gamma^{1,2}_+,\Gamma^{1,2}_-\}$, defined as $\Gamma^{i,j}_\pm = \Gamma_i \pm i \Gamma_j$, transform exactly the same way, and using Table~\subref*{table:coupling33} we can therefore combine them as follows:
\begin{align}
\Gamma^{(1)}: \psi^{(1)} &= \frac{1}{\sqrt{2}}i k_z k_-^2 \Gamma^{1,2}_-  - \frac{1}{\sqrt{2}} ik_z k_+^2 \Gamma^{1,2}_+\nonumber\\
 &\propto k_z(ik_-^2\Gamma^{1,2}_- - ik_+^2\Gamma^{1,2}_+).
\end{align}
This forms an invariant, and thus it is an allowed term in the Hamiltonian. Any multiple of this term is of course also an invariant, and we get a free parameter in front, which we denote $Z_3$. Proceeding this way we achieve the following Hamiltonian, for now excluding electric and magnetic fields:
\begin{align}
H_m = \mathcal E_0 &+ \begin{pmatrix}
\mathcal M & \beta^*_i k_i & 0 & \alpha^*_i k_i \\
\beta_i k_i & -\mathcal M & \alpha^*_i k_i & 0 \\
0 & \alpha_i k_i & \mathcal M & - \beta_i k_i \\
\alpha_i k_i & 0 & -\beta^*_i k_i & -\mathcal M
\end{pmatrix}
\label{modelhamiltonian}\\
&- Z_1\Re(k_+^3)\Gamma_4 + Z_2\Im(k_+^3)\Gamma_3 \nonumber\\
 &+ Z_3k_z(ik_-^2\Gamma^{1,2}_- - ik_+^2\Gamma^{1,2}_+)  \nonumber,
\end{align}
where repeated indices are summed over and where:
\begin{subequations}
\begin{align}
 \mathcal E_0 &=  C + C_{1} \epsilon_{zz} + C_{2}  \epsilon_{||} + D_1 k_z^2 + D_2 k_{||}^2, \\
\mathcal M &= M + M_1 \epsilon_{zz} + M_2 \epsilon_{||}- B_1 k_z^2 - B_2 k_{||}^2, \\
\alpha_1 &=  A_2 + A_{21}\epsilon_{zz} + A_{22}\epsilon_{||} +  A_{23}k_z^2 + A_{24}k_{||}^2  \nonumber\\
&  + i Y_3 \epsilon_{z-} + Y_4\epsilon_+,\\
\alpha_2 &= i A_2 + iA_{21}\epsilon_{zz} + iA_{22}\epsilon_{||} + iA_{23}k_z^2 + iA_{24}k_{||}^2 \nonumber\\
&+  Y_3 \epsilon_{z-} - i Y_4\epsilon_+, \\
\alpha_3 &= Y_1\epsilon_{z+} + i Y_2 \epsilon_-,\\
\beta_1 &= X_1 \epsilon_{zx} +  2X_2\epsilon_{xy} - iX_3  \epsilon_{zy} + iX_4(\epsilon_{xx}-\epsilon_{yy}),\\
\beta_2 &= X_1\epsilon_{zy} + X_2(\epsilon_{xx}-\epsilon_{yy}) + i X_3 \epsilon_{zx} - 2i X_4 \epsilon_{xy}, \\
\beta_3 &=  A_1 + A_{11}\epsilon_{zz} + A_{12}\epsilon_{||} + A_{13}k_z^2 + A_{14}k_{||}^2,
\end{align}
\end{subequations}
with $\epsilon_{||} = \epsilon_{xx} +\epsilon_{yy}$, $\epsilon_{\pm} = \epsilon_{xx} - \epsilon_{yy} \pm 2i\epsilon_{xy}$, $\epsilon_{z\pm} = \epsilon_{zx} \pm i\epsilon_{zy}$. From the method of invariants it follows that all model parameters designated by capital roman letters must be real parameters, which are not determined by the symmetries of the crystal. The Hamiltonian above contains all allowed terms to third order in $\mathbf k$, first order in strain and terms first order in both. Neglecting strain, this Hamiltonian reduces to the  one derived in Ref.~\onlinecite{model}, except for three terms combining the in-plane and out-of-plane components of the wave vector, $k_z^2(k_+\Gamma^{1,2}_- + k_-\Gamma^{1,2}_+)$, $ k_{||}^2 k_z \Gamma_3$ and $k_z(ik_-^2\Gamma^{1,2}_- - ik_+^2\Gamma^{1,2}_+)$, which were not included in Ref.~\onlinecite{model}. The $\mathbf k$ independent strain terms depend only on $\epsilon_{||}$ and $\epsilon_{zz}$, and are either proportional to the identity matrix or $\Gamma_5$. Hence, the effects of these terms can be described within the unstrained model by making the  parameters strain dependent $M \rightarrow \tilde M(\mathbf \epsilon)$ and $C \rightarrow \tilde C(\mathbf \epsilon)$, where:
\begin{subequations}
\begin{align}
  \tilde M(\mathbf \epsilon) &=  M + M_1 \epsilon_{zz} + M_2 \epsilon_{||},\\
 \tilde C(\mathbf \epsilon) &=  C + C_1 \epsilon_{zz} + C_2 \epsilon_{||}.
\end{align}
\end{subequations}
Similarly the terms linear in $\mathbf k$ and $\epsilon_{||}$ or $\epsilon_{zz}$, can be described by strain-dependent model parameters in the unstrained model.  The shear strain terms and the terms with $\epsilon_{xx}-\epsilon_{yy}$ give new terms linear in $\mathbf k$, which cannot be described by making the parameters of the unstrained model strain-dependent. The effects of $\epsilon_{zx}$ are similar to the effects of $\epsilon_{xy}$, since they contribute to the real parts of $\beta_1$, $\alpha_2$ and $\alpha_3$ and the imaginary parts of $\alpha_1$ and $\beta_2$. For the same reason $\epsilon_{xx}-\epsilon_{yy}$ and $\epsilon_{zy}$ have similar effects. As we demonstrate below, the new terms linear in the wave vector give rise to new physics at the $(111)$ surface of a semi-infinite topological insulator.

Higher order terms in strain will always be both inversion and time-reversal symmetric, and in Table~\ref{table:transformation} we see that the only matrices symmetric under both I and TR are the identity matrix and $\Gamma_5$, and $k$-independent strain terms may therefore be lumped into the strain dependent parameters $\tilde M(\mathbf \epsilon)$ and $\tilde C(\mathbf \epsilon)$.

\subsection{Magnetic field}
The Hamiltonian in Eq.~\eqref{modelhamiltonian} is invariant under both TR and inversion, thus all bands are doubly degenerate. The TR symmetry can be broken by a magnetic field, which must be included in the following form:
\begin{align}
H_B &= \frac{\mu_B}{2}\begin{pmatrix}
g_{1z}B_z & 0 & g_{1p}B_- & 0 \\
0 & g_{2z}B_z & 0 & g_{2p}B_- \\
g_{1p}B_+ & 0 & -g_{1z}B_z & 0 \\
0 & g_{2p}B_+ & 0 & -g_{2z}B_z
\end{pmatrix},
\end{align}
where $g_{1p,2p,1z,2z}$ are real parameters and $B_\pm = B_x \pm iB_y$. This contribution was discussed in Ref.~\onlinecite{model}.

\subsection{Electric field}
Inversion symmetry can be broken by an electric field, giving rise to a Hamiltonian of the form:
\begin{align}
H_E &= W_{||}(iE_-\Gamma^{25,51}_+ - i E_+ \Gamma^{25,51}_-  ) + W_z E_z \Gamma_{35}\nonumber\\
&= \begin{pmatrix}
0 & iW_zE_z & 0 & iW_{||}E_- \\
-i W_z E_z & 0 & -iW_{||}E_- & 0 \\
0 & iW_{||}E_+ & 0 & -iW_zE_z \\
-iW_{||}E_+ & 0 & iW_zE_z & 0
\end{pmatrix},
\label{HamiltonianEfield}
\end{align}
where $W_{||,z}$ are real parameters and $E_\pm = E_x \pm iE_y$.

\section{\label{bulk}Bulk spectrum and band-gap}
\begin{figure}
\includegraphics[width=0.43\textwidth]{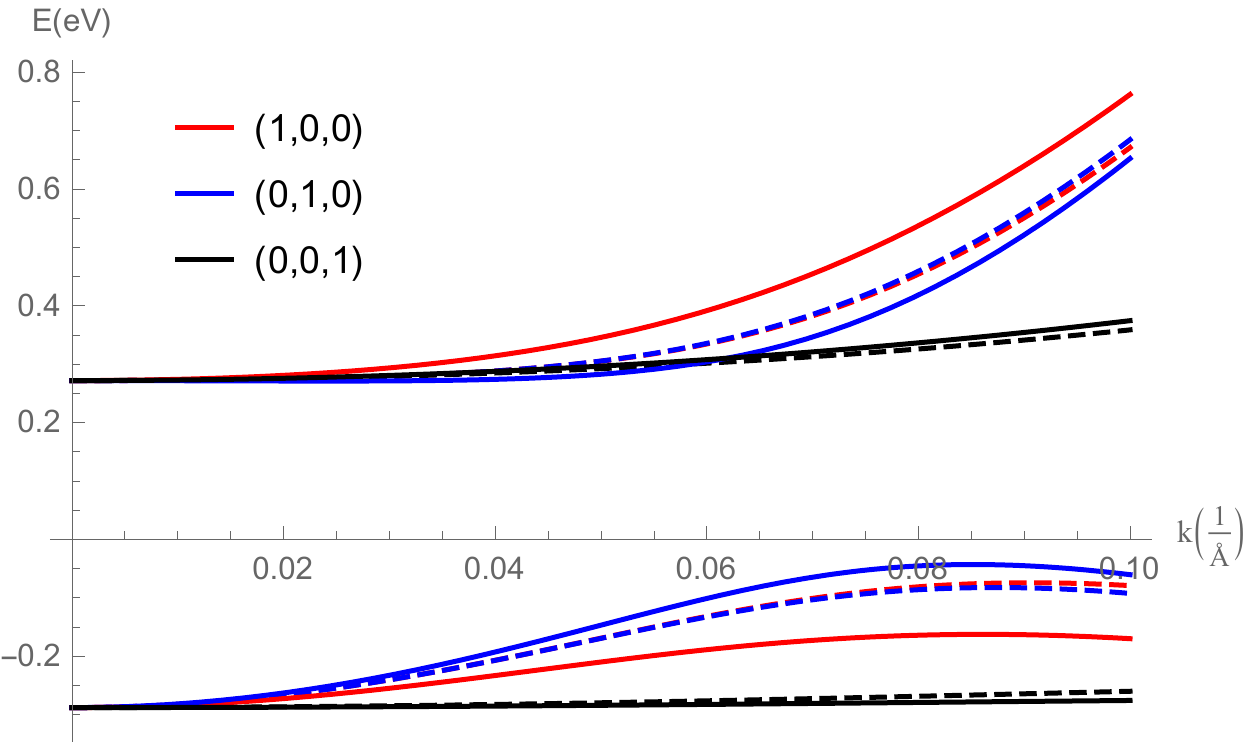}
    \caption{Band structure close to the gamma point, with $\epsilon_{zy}=10\%$ (solid) and without strain (dashed) in the absence of electric and magnetic fields. Without strain the in-plane rotational symmetry is broken only by terms to third order in the wave vector, and only a small splitting between different in-plane direction is seen. Including strain allows terms to first order in the wave vector that breaks the in-plane rotational symmetry, and we see that a splitting of the in-plane directions occurs at lower $k$ values.  Here we have used the parameters of Ref.~\onlinecite{model} for the unstrained model, and $X_i=Y_i=\SI{10}{\electronvolt\angstrom}$. We note that $\epsilon_{xx}-\epsilon_{yy}=10 \%$ gives the same dispersion for the parameters used here.}
    \label{fig:bulkbandstructure}
\end{figure}

\begin{figure}
\centering
Band gap at $\Gamma$ point ($\si{\electronvolt}$)
\includegraphics[width=0.5\textwidth]{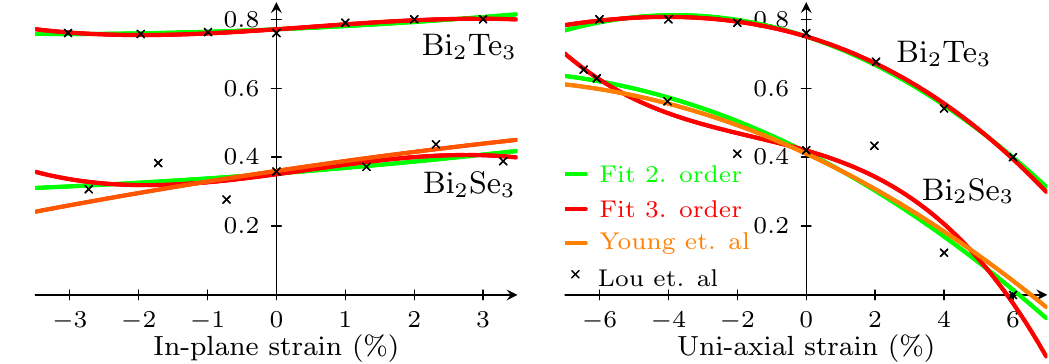}
\caption{The band gap at $\Gamma$ point as a function of strain according to the DFT calculations of Refs.~\onlinecite{young,luo}. The curves from Ref.~\onlinecite{young} are calculated using the approximate linear relationships between the in-plane and out-of-plane lattice constant reported in Ref.~\onlinecite{luo}. The fits to the data from Ref.~\onlinecite{luo} show good agreement, especially for $\text{Bi}_2\text{Te}_3$. Going to third order makes only a slight improvement compared to second order.}
\label{fig:DFTresults}
\end{figure}

In this section we will analyze the effects of strain and electric field on the bulk band structure close to the $\Gamma$ point. The Hamiltonian $H=H_m+H_E$ of Eq.~\eqref{modelhamiltonian} and \eqref{HamiltonianEfield} can be diagonalized analytically giving:
 \begin{align}
E&=\mathcal E_0  \pm \Bigg( \mathcal M^2 + \abs{\alpha_i k_i + iZ_3k_zk_-^2}^2 \nonumber\\
&+ \left|\beta_i k_i -i Z_1\Re(k_+^3) + Z_2\Im(k_+^3) \right|^2 + W_z^2 E_z^2 + W_{||}^2 E_{||}^2 \nonumber\\
&\pm \bigg(2\Big|\left(\alpha_i k_i + iZ_3k_zk_-^2\right)W_{||}E_+ \nonumber\\
&+ \left(\beta_i k_i -iZ_1\Re(k_+^3) + Z_2\Im(k_+^3)\right)W_z E_z\Big|^2 \nonumber\\
&+ 4W_{||}^2 E_{||}^2  \abs{\beta_i k_i -iZ_1\Re(k_+^3) + Z_2\Im(k_+^3) }^2 \nonumber\\
&+ 4 W_z^2 E_z^2  \abs{\alpha_i k_i + iZ_3k_zk_-^2}^2 \nonumber\\
&- 2\Re\Big(\left(\alpha_i k_i + iZ_3k_zk_-^2\right)W_{||}E_+ \nonumber\\
&- \left(\beta_i k_i -iZ_1\Re(k_+^3) + Z_2\Im(k_+^3)\right)W_z E_z\Big)
\bigg)^{\frac{1}{2}} \Bigg)^{\frac{1}{2}}.
\label{bulk_spectrum}
\end{align}
In Fig.~\ref{fig:bulkbandstructure} an example of the band structure with and without strain is plotted. The terms in the Hamiltonian linear in both strain and wave vector split the bands in the $x$ and $y$ directions closer to the $\Gamma$ point compared to the case without strain, where the bands in the $x$ and $y$ directions are only split by the $\mathbf k^3$ terms. The electric field breaks the inversion symmetry, and the degeneracies of the bulk conduction and valence bands are split. At the $\Gamma$ point the degeneracies are protected by TR symmetry which is not broken. Another important quantity of the electronic structure is the gap between the valence and conduction band at the $\Gamma$ point, given by:
\begin{align}
\Delta_{\Gamma} = 2 \sqrt{(M+M_1\epsilon_{zz}+M_2 \epsilon_{||})^2 + W_z^2 E_z^2 + W_{||}^2 E_{||}^2 }.
\end{align}
Note that the gap can only be increased by applying an electric field, whereas strain can increase or decrease the gap depending on the sign of the strain. The effects of strain have been investigated earlier by DFT calculations. In Ref.~\onlinecite{luo} both $\text{Bi}_2\text{Se}_3$ and $\text{Bi}_2\text{Te}_3$ were investigated. Here the lattice constant in the $xy$-plane or the $z$-direction was fixed relative to the known unstrained lattice constant, and the lattice constant in the other direction was relaxed before the calculations were performed. An approximate linear relationship between the in-plane, and out-of-plane lattice constants was reported~\cite{luo}. In Ref.~\onlinecite{young} the band gap was calculated for many different strain configurations and then fitted to a second order polynomial of the strain components. In figure \ref{fig:DFTresults} the results of these calculations are shown together with second and third order fits to the results of Ref.~\onlinecite{luo}. The results show good agreement and both predict a topological phase transition at approximately $6\%$ uni-axial strain. According to our model the topological phase transition will occur when:
\begin{align}
\epsilon_{zz}= -\frac{M}{M_1} - \frac{M_2}{M_1}\epsilon_{||}.
\end{align}

It should be emphasized that strain can only change the gap by a term proportional to the matrix $\Gamma_5$ in the Hamiltonian, which is equivalent to simply changing the band gap parameter $M$. In the following sections we will then simply use a strain-dependent band gap parameter $\tilde M(\mathbf \epsilon)$, and use the parameters of Ref.~\onlinecite{young} to get a quantitative strain dependence.

In general, the parameters of this symmetry-adapted model can be determined by using DFT calculations. The values of the parameters are highly dependent on the computational details as can be seen by comparing Refs.~\onlinecite{oldmodel,model,newmodel}. In this paper,we use the values from Ref.~\onlinecite{model} for the parameters not related to strain.

\subsection{Effective masses}
An important quantity for transport and optical properties of a solid is the effective mass tensor. Here we calculate the diagonal components of the inverse effective mass tensor including both strain and electric field. The inverse effective mass tensor is given by:
\begin{align}
\left[\frac{1}{m^*}\right]_{ij} = \frac{1}{\hbar^2}\frac{\partial^2 E}{\partial k_i \partial k_j}.
\end{align}
With $\mathbf E =0$ we get the effective masses along the $x$ and $y$ axes:
\begin{align}
\left[\frac{\hbar^2}{m}\right]_{ii} &= 2 D_2 \pm \bigg(\frac{|\alpha_i|^2+|\beta_i|^2-2(M+M_1\epsilon_{zz}+M_2 \epsilon_{||})B_2}{\frac{\Delta_\Gamma}{2}}\bigg),
\end{align}
for $i=x,y$ and along the $z$ axis
\begin{align}
\left[\frac{\hbar^2}{m}\right]_{zz} &= 2 D_1 \pm \bigg(\frac{|\alpha_3|^2+|\beta_3|^2 - 2(M+M_1\epsilon_{zz}+M_2 \epsilon_{||})B_1}{\frac{\Delta_\Gamma}{2}}\bigg).
\end{align}
The plus (minus) sign corresponds to the conduction (valence) band. For a non-zero electric field each of the two bands is split into two subbands having the same effective mass, but differing by linear and cubic terms in the wave vector.  The effective masses including both strain and electric field are given in appendix \ref{eff_mass}.

\begin{figure}
{\includegraphics[width=\columnwidth]{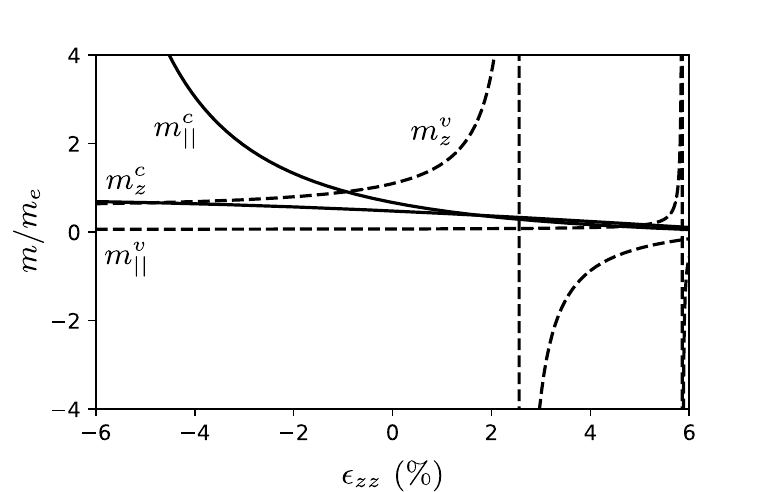}}
\caption{The bulk conduction and valence band effective masses of Bi$_2$Se$_3$ plotted as a function of the strain component $\epsilon_{zz}$ and in the absence of electric and magnetic fields. In the calculations, we have used the DFT computed strain parameters from Ref.~\onlinecite{luo} and Ref.~\onlinecite{model} for the parameters not related to strain. The solid (dashed) lines refer to the conduction (valence) band effective masses.}
\label{fig:effectivemass}
\end{figure}

In Fig.~\ref{fig:effectivemass} the bulk effective masses of Bi$_2$Se$_3$ are plotted as a function of the strain component $\epsilon_{zz}$ and in the absence of electric and magnetic fields. Conduction (valence) band masses are shown as solid (dashed) lines and the effective mass superscript $c$ ($v$) refers to the conduction (valence) band. In this case where only $\epsilon_{zz}\neq 0$ and the electric field is zero the effective masses along the $x$ and $y$ directions are the same. Note that since the maximum of the valence band shifts away from the $\Gamma$ point at a sufficiently large strain value, the valence band effective masses change as a function of strain from positive to negative and reach infinite values in-between.

\section{\label{semiinfinite} Surface state spectrum and spin structure}
\begin{figure}
    \centering
 \subfloat[Group velocity \label{fig:fermivelocity}]{
 \begin{tikzpicture}
\node at (0,0) {\includegraphics[width=0.19\textwidth]{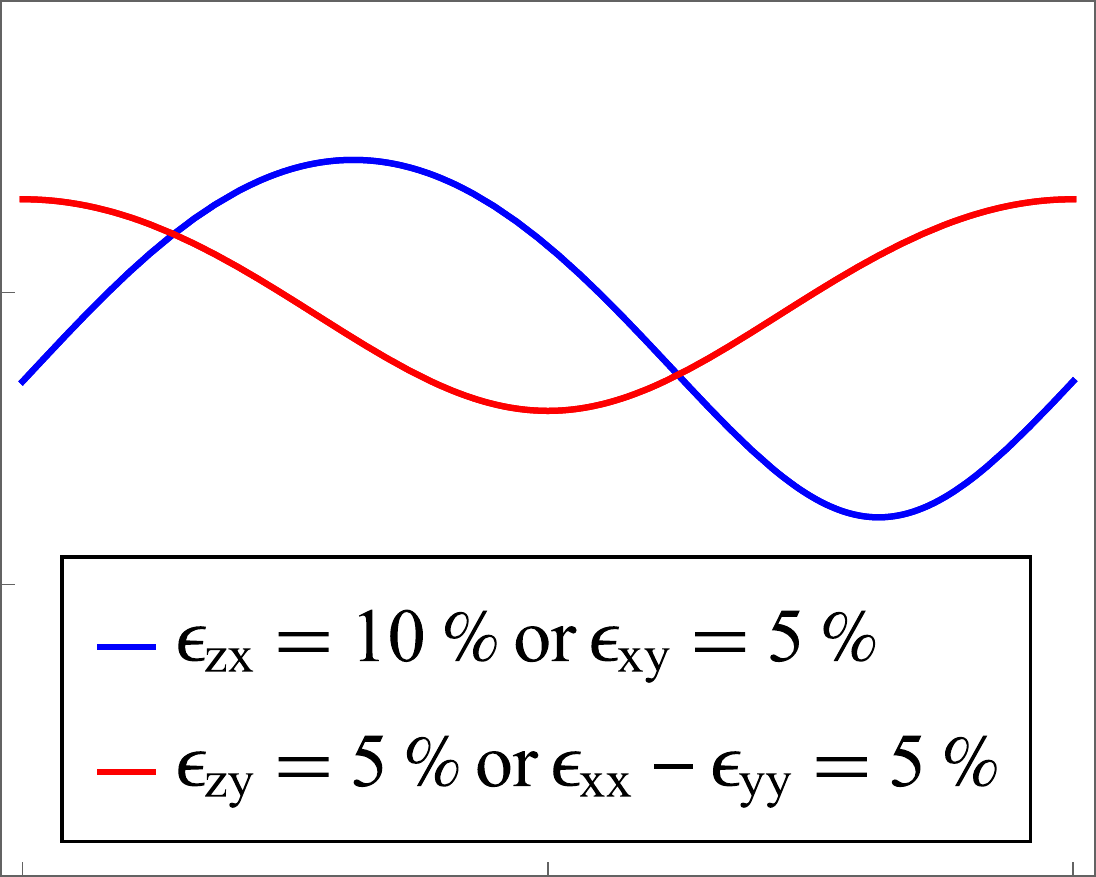}};
\node[rotate=90] at (-2,0) {\footnotesize{$v_F/v_F^0$}};
\node at (-1.78,0.45) {\scriptsize{1}};
\node at (-1.8,-1.35) {\scriptsize{0}};
\node at (0,-1.55) {\scriptsize{$\frac{\pi}{2}$}};
\node at (-1.62,-1.5) {\scriptsize{0}};
\node at (1.62,-1.5) {\scriptsize{$\pi$}};
\node at (0,-1.9) {\footnotesize{$\theta$}};
\end{tikzpicture}
}
\subfloat[Unstrained $\text{Bi}_2\text{Se}_3$ \label{fig:spin_nostrain}]{
\begin{tikzpicture}
\node[inner sep=0pt] at (0,0)
            {\includegraphics[width=0.235\textwidth]{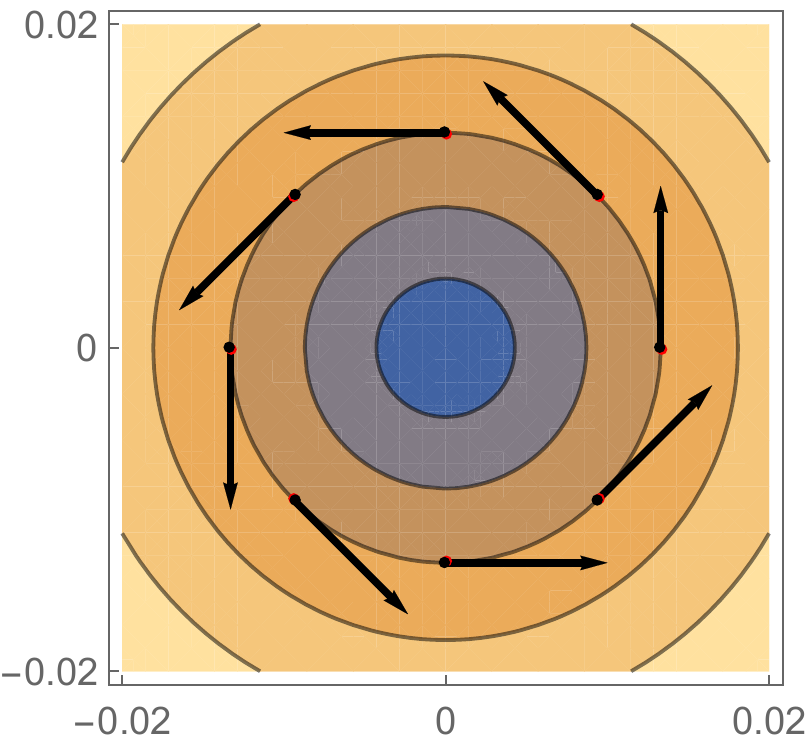}};
            \node[] at (0.3,-2) {\footnotesize{$k_x \text{ (\si{\per\angstrom})} $}};
\node[rotate=90] at (-1.9,0.2) {\footnotesize{$k_y \text{ (\si{\per\angstrom})} $}};
            \end{tikzpicture}}\\
\subfloat[$\epsilon_{zx}=10\%$ \label{fig:spin_strainzx}]{
\begin{tikzpicture}
\node[inner sep=0pt] at (0,0)
            {\includegraphics[width=0.235\textwidth]{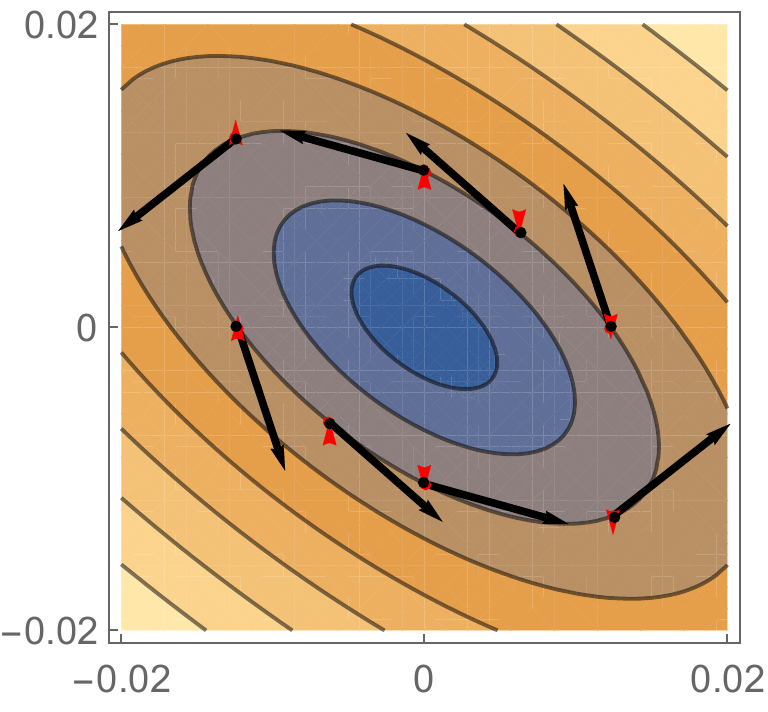}};
            \node[] at (0.3,-2) {\footnotesize{$k_x \text{ (\si{\per\angstrom})} $}};
\node[rotate=90] at (-1.9,0.2) {\footnotesize{$k_y \text{ (\si{\per\angstrom})} $}};
            \end{tikzpicture}}
\subfloat[$\epsilon_{zy}=5\%$ \label{fig:spin_strainzy}]{
\begin{tikzpicture}
\node[inner sep=0pt] at (0,0)
            {\includegraphics[width=0.235\textwidth]{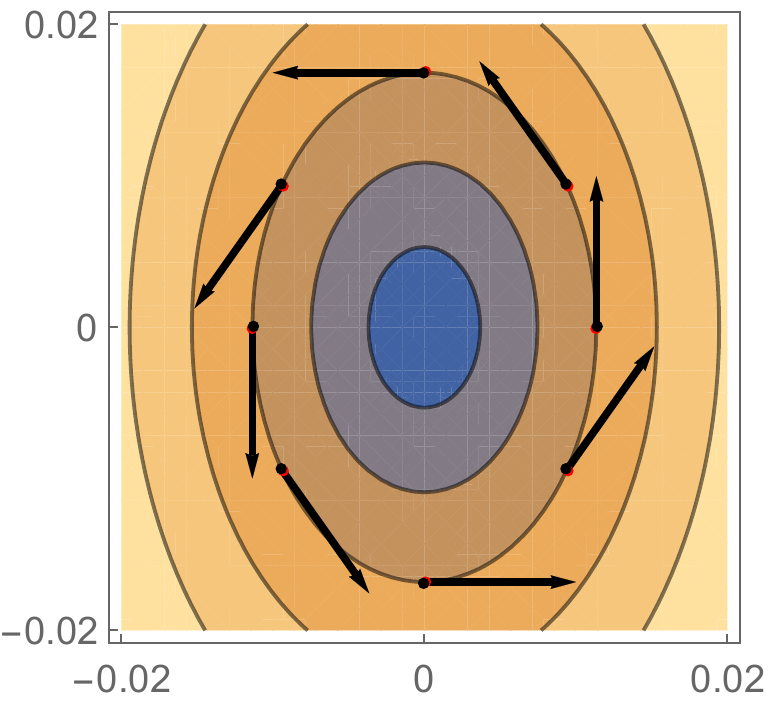}};
            \node[] at (0.3,-2) {\footnotesize{$k_x \text{ (\si{\per\angstrom})} $}};
\node[rotate=90] at (-1.9,0.2) {\footnotesize{$k_y \text{ (\si{\per\angstrom})} $}};
            \end{tikzpicture}}\\
\subfloat[$\epsilon_{xy}=5\%$ \label{fig:spin_strainxy}]{
\begin{tikzpicture}
\node[inner sep=0pt] at (0,0)
            {\includegraphics[width=0.235\textwidth]{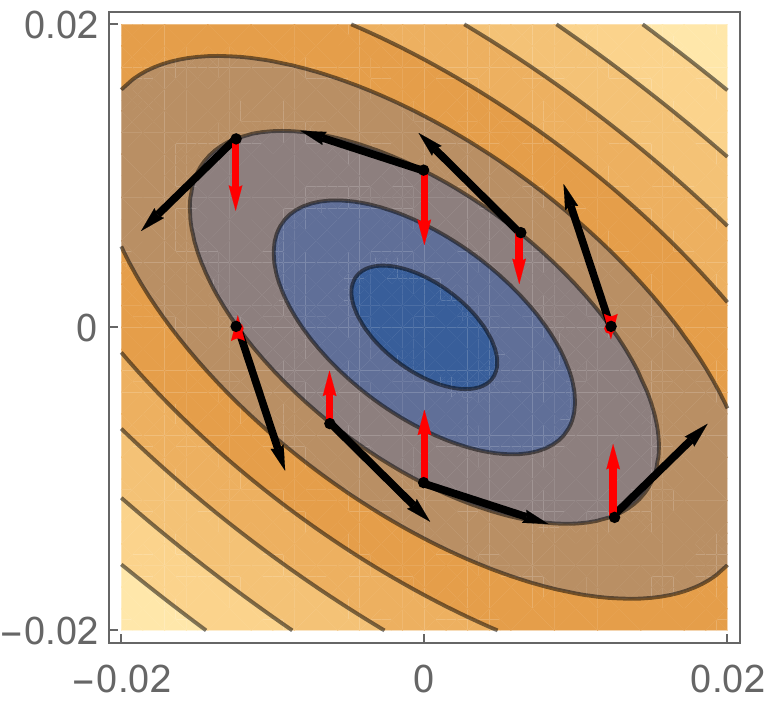}};
            \node[] at (0.3,-2) {\footnotesize{$k_x \text{ (\si{\per\angstrom})} $}};
\node[rotate=90] at (-1.9,0.2) {\footnotesize{$k_y \text{ (\si{\per\angstrom})} $}};
            \end{tikzpicture}}
\subfloat[$\epsilon_{xx} - \epsilon_{yy}=5\%$ \label{fig:spin_strainxxmyy}]{
\begin{tikzpicture}
\node[inner sep=0pt] at (0,0)
            {\includegraphics[width=0.235\textwidth]{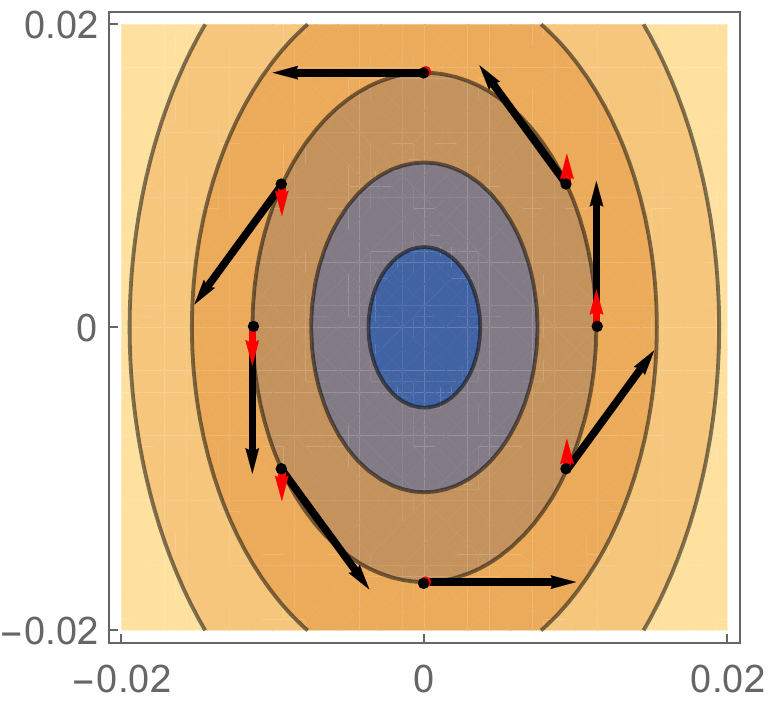}};
            \node[] at (0.3,-2) {\footnotesize{$k_x \text{ (\si{\per\angstrom})} $}};
\node[rotate=90] at (-1.9,0.2) {\footnotesize{$k_y \text{ (\si{\per\angstrom})} $}};
            \end{tikzpicture}}
    \caption{\protect\subref{fig:fermivelocity} Group velocity relative to the velocity in unstrained $\text{Bi}_2\text{Se}_3$ given by $v_F^0=A_2\sqrt{1-\frac{D_1^2}{B_1^2}}$ as a function of the angle $\theta=\arctan(\frac{k_y}{k_x})$ in the $k_x,k_y$ plane. (b)-(f) Contour plots showing the elliptical curves of constant energy of the upper Dirac cone of the surface states in the $(k_x,k_y)$ plane. The black arrows indicate the direction in-plane part of the expectation value of the spin, while the red arrows show the expectation value of the spin in the $z$ direction. Spin-momentum locking is still present, but spin and momentum are no longer perpendicular. Here we have used the parameters of Ref. \onlinecite{model} for the parameters of the unstrained model, and $X_i=Y_i=\SI{10}{\electronvolt\angstrom}$.}
    \label{fig:spinstructure}
\end{figure}

We shall now proceed to calculate the surface state (SS) spectrum and spin structure for the model Hamiltonian \eqref{modelhamiltonian}. In this and the following section we exclude $\mathbf k^3$ terms, i.e. the last three terms in Eq.~\eqref{modelhamiltonian} and the $\mathbf k$ dependent terms in $\alpha_i$ and $\beta_i$. First, we consider a semi-infinite topological insulator filling the $z<0$ half space, implemented via the boundary conditions $\Psi(0)=0$ and $\Psi(z)\rightarrow 0$ for $z\rightarrow -\infty$. The translational invariance is broken in the $z$ direction and we make the substitution $k_z\rightarrow-i\partial_z$. Using the ansatz $\psi_\lambda e^{z\lambda}$ the stationary Schrödinger equation implies the following secular equation:
\begin{align}
 \Leftrightarrow 0 &= D_+D_- \lambda^4 +\bigg( 2(E - \tilde C(\mathbf \epsilon) - D_2 k_{||}^2)D_1 \nonumber\\
 & - 2(\tilde M(\mathbf \epsilon) - B_2 k_{||}^2)B_1 + |\alpha_3|^2 + |\beta_3|^2 \bigg) \lambda^2 \nonumber\\
  & + i(2\Re(\alpha_1\alpha_3^* + \beta_1\beta_3^*) k_x + 2\Re(\alpha_2\alpha_3^* + \beta_2\beta_3^*) k_y)\lambda \nonumber\\
  &+ (E-\tilde C(\mathbf \epsilon) - D_2 k_{||}^2)^2 - (\tilde M(\mathbf \epsilon) -B_2 k_{||}^2)^2\nonumber\\
  & - |\alpha_1 k_x + \alpha_2 k_y |^2
  - |\beta_1 k_x + \beta_2 k_y |^2,
  \label{secularequation}
\end{align}
where $D_\pm=D_1 \pm B_1$. This equation has 4 complex solutions $\lambda_i$, each with two corresponding eigenspinors $\psi_j(\lambda_i)$ given by:
\begin{align}
\psi_1(\lambda_i) &= \begin{pmatrix}
E - \mathcal E_0 + \mathcal M \\
\beta_1 k_x + \beta_2 k_y - i \beta_3 \lambda_i \\
0 \\
\alpha_1 k_x + \alpha_2 k_y - i \alpha_3 \lambda_i
\end{pmatrix},\\
\psi_2(\lambda_i) &= \begin{pmatrix}
\beta_1^* k_x + \beta_2^* k_y - i \beta_3^* \lambda_i \\
E - \mathcal E_0 - \mathcal M \\
\alpha_1 k_x + \alpha_2 k_y - i \alpha_3 \lambda_i \\
0
\end{pmatrix}.
\end{align}
The general solution to the Schrödinger equation can then be written on the form:
\begin{align}
\Psi(z) = \sum_{j=1}^2 \sum_{i=1}^4 C_{ij} \psi_j(\lambda_i)e^{\lambda_i z},
\label{generalwavefunction}
\end{align}
where the coefficients $C_{ij}$ are to be determined from the boundary conditions. For surface states we require that $\Psi(z) \rightarrow 0$ for $z\rightarrow -\infty$, implying that $\Re(\lambda_i)>0$. To satisfy the boundary condition at the surface we need two different solutions with positive real part. By complex conjugation of Eq.~\eqref{secularequation} we see that for a solution $\lambda_i$, $-\lambda_i^*$ will also be a solution. Hence, the solutions are either imaginary or occur in pairs related by $\lambda_i = - \lambda_j^*$. This allows for three distinct cases: (i) All solutions are imaginary. (ii) Two complex solutions related by $\lambda_1 =-\lambda_2^*$ and two imaginary solutions. (iii) Two pairs of complex solutions $\lambda_1=-\lambda_3^*$ and $\lambda_2=-\lambda_4^*$. Existence of SSs is only possible in case (iii) and we therefore assume that $\lambda_1$ and $\lambda_2$ have positive real parts. Hence, we limit the sum in Eq.~\eqref{generalwavefunction} to $i\in\{1,2\}$. Applying the boundary condition at $z=0$ we obtain the following secular equation for nontrivial solution to the coefficients $C_{ij}$:
\begin{align}
(\lambda_1 + \lambda_2)^2 = \frac{|\alpha_3|^2 + |\beta_3|^2}{-D_+D_-},
\label{lambdasum}
\end{align}
and combining this with Eq.~\eqref{secularequation} we find the SS spectrum:
\begin{align}
E &= \tilde C(\mathbf \epsilon) + \frac{D_1}{B_1}\tilde M(\mathbf \epsilon) \nonumber\\
&\pm\sqrt{1-\frac{D_1^2}{B_1^2}}\Big(|\alpha_1 k_x + \alpha_2 k_y |^2 + |\beta_1 k_x + \beta_2 k_y |^2\nonumber\\
&- \frac{\Re((\alpha_1 k_x + \alpha_2 k_y)\alpha_3^* + (\beta_1 k_x +  \beta_2 k_y)\beta_3^*)^2}{|\alpha_3|^2+ |\beta_3|^2}\Big)^{\frac{1}{2}} \nonumber\\
&+ \left(D_2 - \frac{D_1}{B_1} B_2\right)k_{||}^2.
\label{ss_spectrum_3D}
\end{align}
For small $k_{||}$ we see that the spectrum is linear but due to the strain, the group velocity has now become anisotropic and the contours of constant energy have become elliptical. Note that the position of the Dirac-point can be shifted by changing $\tilde M(\mathbf \epsilon)$, however the relative position within the bulk band gap is not changed. Notice also that the second order term is not changed by strain, since we have systematically retained terms of order $\epsilon^{1}k^{1}$ and discarded terms of order $\epsilon^{1}k^{2}$. Both the directional dependence of the group velocity and the ellipticity of the contours of constant energy are shown in Fig.~\ref{fig:spinstructure}.

For $k_{||}=0$, the linear term in Eq.~\eqref{secularequation} vanishes and $\lambda_1$ and $\lambda_2$ can be calculated analytically:
\begin{align}
\lambda_\alpha =   \sqrt{\frac{-F + (-1)^\alpha\sqrt{R}}{2D_+D_-}},
\label{gammapoint_lambda}
\end{align}
where:
\begin{subequations}
\begin{align}
F&=|\beta_3|^2 + |\alpha_3|^2 - 2\tilde M(\mathbf \epsilon) B_1 + 2ED_1 - 2\tilde C(\mathbf \epsilon)D_1, \\
R&=F^2 - 4D_+D_-(E-\tilde C(\mathbf \epsilon) + \tilde  M(\mathbf \epsilon))(E- \tilde C(\mathbf \epsilon) -\tilde M(\mathbf \epsilon)).
\end{align}
\end{subequations}
Imposing the boundary condition $\Psi(0)=0$ with Eq.~\eqref{generalwavefunction} we find the exact wave functions of the two degenerate eigenstates at $k_{||}=0$, which are related by TR:
\begin{subequations}
\label{gammapointwavefunctions}
\begin{align}
\Psi_1 &=N
\begin{pmatrix}
i\sgn(D_+) \sqrt{\frac{|D_+|}{|2B_1|}} \\ \frac{\beta_3}{\sqrt{|\alpha_3|^2+|\beta_3|^2}}\sqrt{\frac{|D_-|}{|2B_1|}} \\ 0 \\ \frac{\alpha_3}{\sqrt{|\alpha_3|^2+|\beta_3|^2}}\sqrt{\frac{|D_-|}{|2B_1|}}
\end{pmatrix}(e^{\lambda_1 z}-e^{\lambda_2 z}), \label{gammapointwavefunctions1}\\
\Psi_2 &= N
\begin{pmatrix}
0 \\ \frac{\alpha_3^*}{\sqrt{|\alpha_3|^2+|\beta_3|^2}}\sqrt{\frac{|D_-|}{|2B_1|}} \\ i\sgn(D_+) \sqrt{\frac{|D_+|}{|2B_1|}} \\  \frac{-\beta_3^*}{\sqrt{|\alpha_3|^2+|\beta_3|^2}}\sqrt{\frac{|D_-|}{|2B_1|}}
\end{pmatrix}(e^{\lambda_1 z}-e^{\lambda_2 z}).
\label{gammapointwavefunctions2}
\end{align}
\end{subequations}
Note that the $\alpha_3$ term couples the spin blocks and in contrast to the case without strain the eigenstates at the $k_{||}=0$ point are mixed up/down spinstates. The conditions for existence of surface states at $k_{||}=0$ are:
\begin{align}
D_+D_- < 0 \quad  \text{and} \quad \tilde M(\mathbf{\epsilon} ) B_1 > 0.
\end{align}
As expected, the surface states can only be destroyed by changing the sign of $\tilde M(\mathbf \epsilon)$, i.e. closing and reopening the bulk band gap.

Due to TR-symmetry the spin of the SS at $(k_x,k_y)$ must be opposite of the spin of the SS at $(-k_x,-k_y)$. For a $(111)$ surface without strain, the model Hamiltonian is invariant under rotations in the $xy$-plane of any angle. Hence $\langle S_z \rangle=0$ since $(k_x,k_y)$ and $(-k_x,-k_y)$ are related by a rotation around the $z$-axis which does not change $S_z$. The terms linear in both wave vector and strain break the full rotation symmetry, and non-zero  $\langle S_z \rangle$ values are allowed. At finite $k_{||}$, Eq.~\eqref{secularequation} was solved numerically, and using $\Psi(0)=0$ a numerical expression for the SS spinor was determined. Finally, the expectation values of the spin operators were calculated and are shown in figure \ref{fig:spinstructure}. As expected, non-zero values of $\langle S_z \rangle$ occur. Such out-of-plane spin components also show up to third order in $k$ in the case of hexagonal warping\cite{warping}. The actual values of $S_{z}$ shown in Fig.~\subref*{fig:spin_strainxy} are of course strongly dependent on the chosen parameters, but the non-zero $S_{z}$-component is generally present for any values of $X{i}$ and $Y_{i}$.

\subsection{\label{2Dmodel} Effective 2D model}
Using the surface states at $k_{||}=0$ as basis we can derive an effective 2D model for the surface states. We use the wave functions in Eq.~\eqref{gammapointwavefunctions}, but with $\epsilon_{ij}=0$ to have a basis independent of strain. We note that in the absence of strain, $\psi_1$ and $\psi_2$ represent spin up/down states respectively. The matrix elements of the  $\Gamma$-matrices in this basis read:
\begin{subequations}
\begin{align}
\bra{\Psi_i} \Gamma_1 \ket{\Psi_j} &= -\sgn(B_1A_1)\sqrt{1-\frac{D_1^2}{B_1^2}}[\sigma_y]_{ij},\\
\bra{\Psi_i} \Gamma_2 \ket{\Psi_j} &= \sgn(B_1A_1)\sqrt{1-\frac{D_1^2}{B_1^2}}[\sigma_x]_{ij},\\
\bra{\Psi_i} \Gamma_3 \ket{\Psi_j} &= 0, \\
\bra{\Psi_i} \Gamma_4 \ket{\Psi_j} &= -\sgn(B_1A_1)\sqrt{1-\frac{D_1^2}{B_1^2}}[\sigma_z]_{ij}, \\
\bra{\Psi_i} \Gamma_5 \ket{\Psi_j} &= \frac{D_1}{B_1}[\sigma_0]_{ij}.
\end{align}
\end{subequations}
The effective 2D model including $\mathbf k$ up to second order becomes:
\begin{align}
H^{2D} =  \tilde C(\mathbf \epsilon) + \frac{D_1}{B_1}\tilde M(\mathbf \epsilon) + \mathbf d(\mathbf k) \cdot \mathbf{\sigma} +  \left(D_2 - \frac{D_1}{B_1} B_2\right)k_{||}^2,
\end{align}
where $\mathbf d(\mathbf k)=(d_1(\mathbf k),d_2(\mathbf k),d_3(\mathbf k))$ is given by:
\begin{subequations}
\begin{align}
d_1(\mathbf k) &= \sqrt{1- \frac{D_1^2}{B_1^2}}\Im(\alpha_1 k_x + \alpha_2 k_y)\\
 &= \sqrt{1- \frac{D_1^2}{B_1^2}} \bigg( \big(Y_3 \epsilon_{zx} + 2 Y_4 \epsilon_{xy}\big) k_x \nonumber\\
&+ \big( A_2 + A_{21}\epsilon_{zz} + A_{22}\epsilon_{||} - Y_3\epsilon_{zy}-Y_4(\epsilon_{xx}-\epsilon_{yy})\big) k_y\bigg), \nonumber\\
d_2(\mathbf k) &=-\sqrt{1- \frac{D_1^2}{B_1^2}}  \Re(\alpha_1 k_x + \alpha_2 k_y)\\
&=- \sqrt{1- \frac{D_1^2}{B_1^2}} \bigg((Y_3\epsilon_{zy} + 2 Y_4 \epsilon_{xy})k_y \nonumber\\
&+   \big(A_2 + A_{21}\epsilon_{zz} + A_{22}\epsilon_{||} + Y_3\epsilon_{zy} + Y_4(\epsilon_{xx}-\epsilon_{yy}) \big) k_x\bigg), \nonumber \\
d_3(\mathbf k) &= - \sqrt{1- \frac{D_1^2}{B_1^2}}\Im(\beta_1 k_x + \beta_2 k_y) \\
&= \sqrt{1- \frac{D_1^2}{B_1^2}}\bigg(\big(X_3\epsilon_{zy} (\epsilon_{xx}-\epsilon_{yy})\big)k_x \nonumber\\
&+ \big(-X_3 \epsilon_{zx} + 2X_4\epsilon_{xy}\big)k_y\bigg). \nonumber
\end{align}
\end{subequations}
Diagonalizing the effective Hamiltonian gives the spectrum:
\begin{align}
E &=  \tilde C(\mathbf \epsilon) + \frac{D_1}{B_1}\tilde M(\mathbf \epsilon) +  \left(D_2 - \frac{D_1}{B_1} B_2\right)k_{||}^2\\
&\pm  \sqrt{1- \frac{D_1^2}{B_1^2}}\sqrt{|\alpha_1 k_x + \alpha_2 k_y|^2 +  \Im(\beta_1 k_x + \beta_2 k_y)^2 }. \nonumber
\end{align}
This spectrum is different from Eq.~\eqref{ss_spectrum_3D}, obtained from the 3D model. The two spectra become equal, however, if we take $\alpha_3=0$. We note that the Berry curvature of the SS band vanishes identically, i.e. $\Omega(\mathbf k) = \hat d \cdot [(\partial_{k_x} \hat d) \times (\partial_{k_y} \hat d)]=0$.

\section{\label{finite} Localization of surface states and finite size effect}
In the case of a finite slab, a gap will open in the SS spectrum at $k_{||}=0$. This is due to the overlap of SS on opposite surfaces. For a given thickness the overlap will be highly dependent on the penetration depth of the SS. From the wave functions the expectation value of the distance from the surface $d=-\langle z \rangle$ can be calculated to:
\begin{align}
d=  \frac{\frac{B_1}{2\tilde M(\mathbf{\epsilon})} + \frac {-D_+D_-}{|\alpha_3|^2+|\beta_3|^2} }{\sqrt{\frac {-D_+D_-}{|\alpha_3|^2+|\beta_3|^2}}},
\end{align}
and we see that $d$ depends on strain through the band gap parameter, and the quantity $\sqrt{|\alpha_3|^2+|\beta_3|^2}$ involving the coefficients of the terms first order in $k_z$.

The eigenenergies in the finite system can be calculated by imposing the boundary condition $\Psi(\pm L/2)=0$, where $L$ is the slab thickness. Now the wave function can be written as Eq.~\eqref{generalwavefunction}, but using all 4 solutions to Eq.~\eqref{secularequation}. At the $\Gamma$ point, the solutions are $\pm \lambda_{1,2}$ with $\lambda_{1,2}$ given by Eq.~\eqref{gammapoint_lambda}. The secular equation of the nontrivial solution for the coefficients gives:
\begin{align}
\frac{(E-\tilde C(\mathbf{\epsilon}) + \tilde M(\mathbf{\epsilon}) + D_+\lambda_1^2)\lambda_2}{(E-\tilde C(\mathbf{\epsilon}) + \tilde M(\mathbf{\epsilon}) + D_+\lambda_2^2)\lambda_1} = \left(\frac{\tanh(\lambda_1 L/2)}{\tanh(\lambda_2 L/2)}\right)^{\pm 1},
\label{equation_finite}
\end{align}
for even/odd states respectively. This is the same form as found using the model without strain \cite{njp_films} , but here the parameters $\tilde M(\mathbf{\epsilon})$ and $\tilde C(\mathbf{\epsilon})$  as well as  $\lambda_{1,2}$ from Eq.~\eqref{gammapoint_lambda} depend on strain. The parameter $\tilde C(\mathbf{\epsilon})$ is not important, since it simply shifts all energies. Again the important quantities are $\tilde M(\mathbf{\epsilon})$ and $\sqrt{|\alpha_3|^2+|\beta_3|^2}$.

\begin{figure}
\subfloat[\label{fig:gap_epsilonzz}]{\begin{tikzpicture}
\node[inner sep=0pt] (russell) at (0,0)
    {\includegraphics[width=0.4\textwidth]{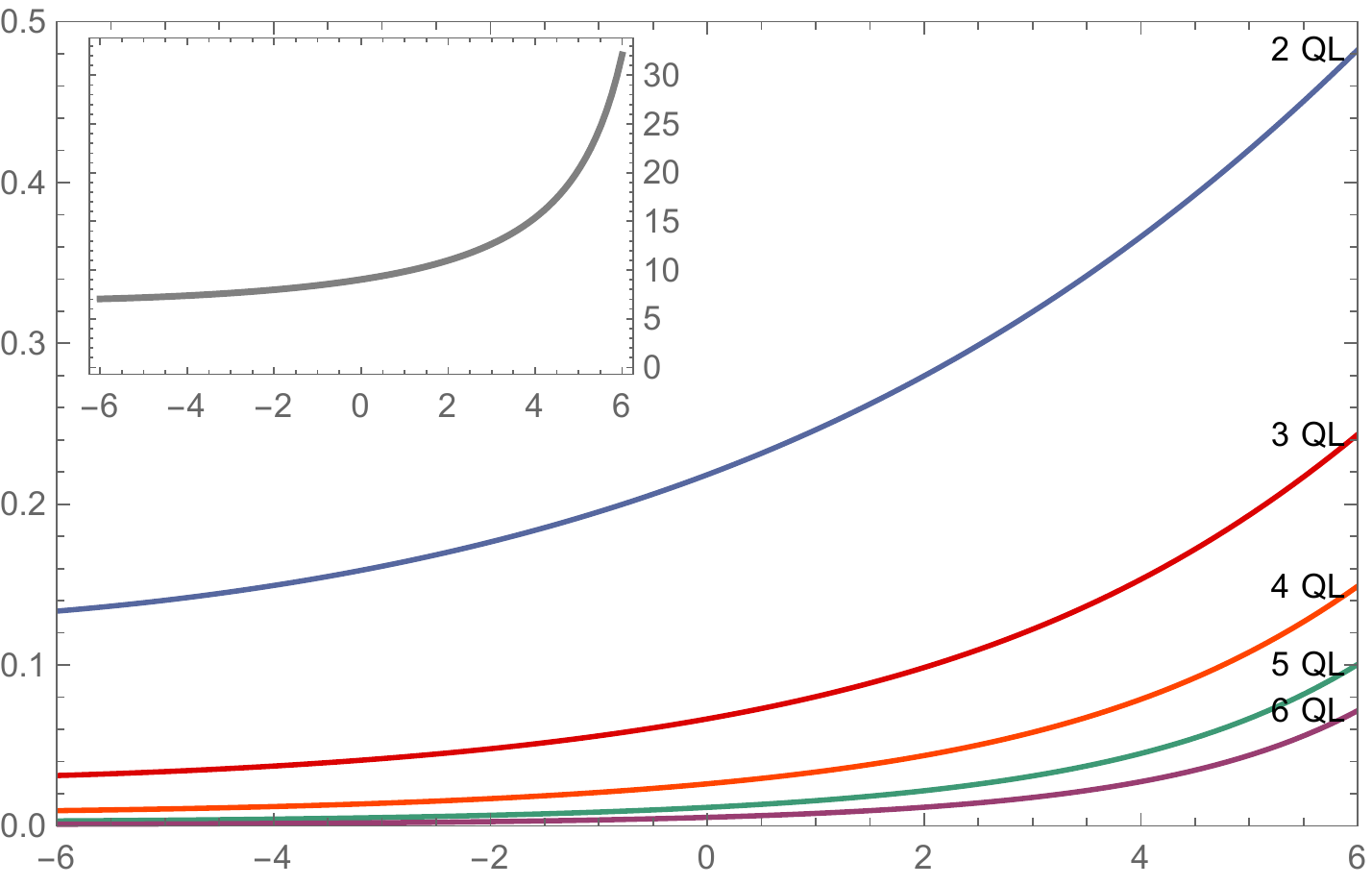}};
\node[] at (0,-2.5) {\footnotesize{$\epsilon_{zz} \text{ (\%)} $}};
\node[rotate=90] at (-3.8,0) {\footnotesize{$\Delta \text{ (eV)} $}};
\node[rotate=90] at (0.2,1) {\footnotesize{$-\langle z \rangle \text{(\si{\angstrom})}$}};
\end{tikzpicture}}\\
\subfloat[\label{fig:gap_alphabeta}]{\begin{tikzpicture}
\node[inner sep=0pt] (russell) at (0,0)
  {\includegraphics[width=0.4\textwidth]{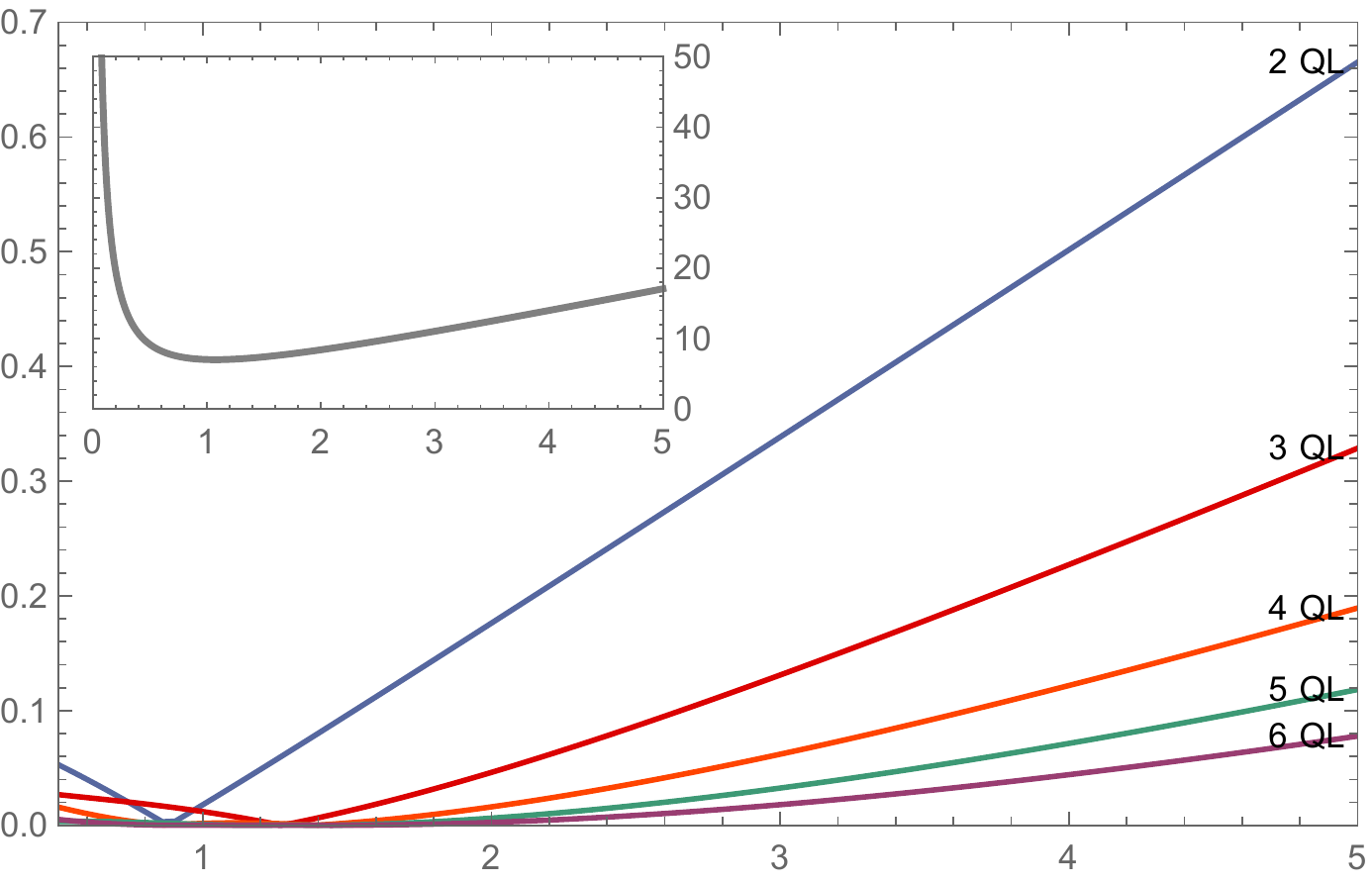}};
\node[] at (0,-2.5) {\footnotesize{$\sqrt{|\alpha_3|^2+|\beta_3|^2} \text{ (\si{\electronvolt\angstrom})} $}};
\node[rotate=90] at (-3.8,0) {\footnotesize{$\Delta \text{ (eV)} $}};
\node[rotate=90] at (0.4,1.0) {\footnotesize{$-\langle z \rangle \text{(\si{\angstrom})}$}};
\end{tikzpicture}}
\label{fig:gap}
\caption{Band gap for thicknesses of 2 to 6 quintuple layers (QL). In \protect\subref{fig:gap_epsilonzz} the effect of changing the bulk band gap by uni-axial strain on the SS gap $\Delta$ is shown. We have used a relation between the bulk band gap and $\epsilon_{zz}$ from Ref.~\onlinecite{young}. In this plot we have neglected the strain dependence of $\sqrt{|\alpha_3|^2+|\beta_3|^2}$ by setting $A_{11} = 0$. In \protect\subref{fig:gap_alphabeta} we show the SS gap as a function of  $\sqrt{|\alpha_3|^2+|\beta_3|^2}$, assuming a constant bulk band gap, i.e., setting $M_1 = M_2 = 0$. The insets shows the expectation value of the distance to the surface for the surface state in the semi-infinite case, with the same horizontal axis.  A qualitative agreement between the localization of the SS on single surface and the gap induced by coupling between opposite surfaces is seen. We have used the parameters of the model without strain from Ref.~\onlinecite{model}.}
\end{figure}
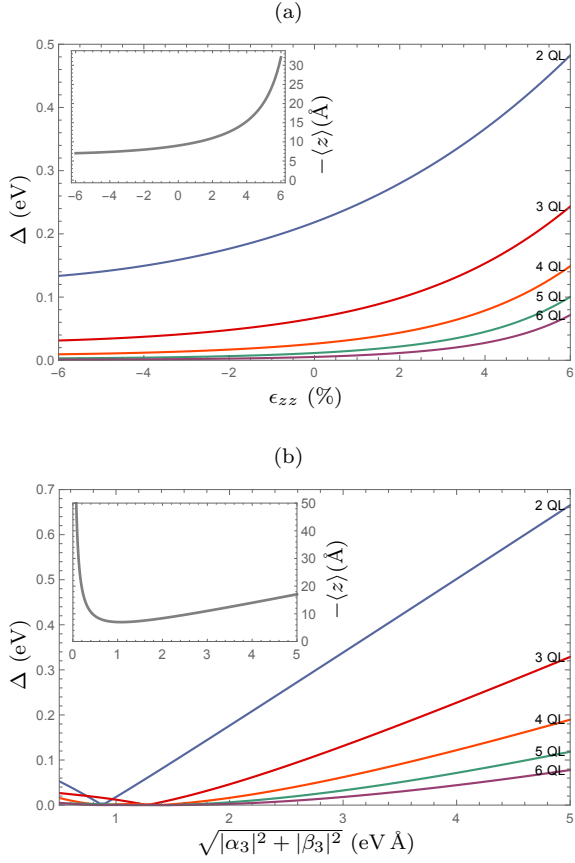

To find the gap of the SSs, $\Delta$, Eq.~\eqref{equation_finite} has been solved numerically for $L=n \cdot \SI{9.547}{\angstrom}$ for integers $n$ between 2 and 6 where $L=\SI{9.547}{\angstrom}$ is the thickness of 1 QL. First we have analyzed the effect of the bulk band gap changed by strain, and taking  $\sqrt{|\alpha_3|^2+|\beta_3|^2}=A_1$ the value without strain. We see that increasing the bulk band gap, decreases the SS band gap. Close to the phase transition this behavior is evident, since the SS must approach bulk states extending further into the material as the bulk gap approaches zero, giving a large coupling between opposite surfaces. As we saw in section \ref{bulk}, the bulk gap is most sensitive to $\epsilon_{zz}$. Using the relation
$$\Delta E_g = (\SI{-5.27}{\electronvolt})\epsilon_{zz} + \tfrac{1}{2}(\SI{-60.1}{\electronvolt})\epsilon_{zz}^2, $$
from Ref.~\onlinecite{young} with $2|\tilde M(\mathbf \epsilon)|=E_g$ we have plotted the SS gap as function of $\epsilon_{zz}$ in Fig.~\ref{fig:gap_epsilonzz}, using the value $M=\SI{0.28}{\electronvolt}$ from Ref.~\onlinecite{model} for the band gap in unstrained $\text{Bi}_2\text{Se}_3$ and setting $A_{11} = 0$.

In Fig.~\ref{fig:gap_alphabeta}, we have plotted the dependence of the SS gap on $\sqrt{|\alpha_3|^2+|\beta_3|^2}$ assuming a constant bulk band gap parameter of $M=\SI{0.28}{\electronvolt}$ from Ref.~\onlinecite{model} and setting $M_1 = M_2 = 0$. Without strain $\sqrt{|\alpha_3|^2+|\beta_3|^2}=A_1=\SI{2.26}{\electronvolt\angstrom}$ according to Ref.~\onlinecite{model}, but the quantitative dependence on strain has not been determined. We see that the SS gap closes at low values, which is related to the fact that for $|\alpha_3|^2+|\beta_3|^2<\frac{4\tilde M |D_+D_-|}{B_1}$ the solutions to Eq.~\eqref{secularequation}, $\lambda_i$, are complex giving an oscillatory spatial dependence of the SS. Oscillatory surface states give oscillations in the SS gap as a function of $L$, which has been theorized to signify topological phase transitions between a trivial 2D insulator and a quantum spin hall phase\cite{osc_gap1,osc_gap2,osc_gap3}. We note here that these oscillations are present only using the parameters of Ref.~\onlinecite{oldmodel}, not the parameters of Ref.~\onlinecite{model} as shown in Ref.~\onlinecite{speciale}. Experimental evidence of oscillations in the gap remains inconclusive, since both a simple decay of the gap for 2-5 QL has been reported\cite{experiment_gap} as well as an increase in the gap from 2 to 3 QL\cite{experimental_gap_osc} signifying a topological phase transition between 2 and 3 QL. The observed gaps are summarized in Table~\ref{table:experimental_gap}.

We have demonstrated that the gap of the SS can be tuned by strain. However, quantitative predictions require access to the parameters relating the strain tensor to  $\sqrt{|\alpha_3|^2+|\beta_3|^2}$. Tuning the band gap of the SS is important for both applications and fundamental research. One particular experimental challenge is that the layered structure of $\text{Bi}_2\text{Se}_3$ makes it easy to fabricate integer numbers of quintuple layers. Thus the gap dependence on the layer thickness has only been experimentally studied with few data points, since the gap becomes too small to be measured already at 6 QL\cite{experiment_gap}. With strain, it is be possible to increase the SS gap and investigate the thickness dependence of the gap further.  Another intriguing prospect would be tuning the gap via strain, so as to pass through the topological transition in a controlled manner.

\begin{table}[]
\centering
\begin{tabular}{|c|c|c|c|c|}
\hline
 $L$ & 2 QL & 3 QL & 4 QL & 5 QL  \\\hline
 $\Delta(\si{\electronvolt})$ & $0.25$  & $0.138$ & $0.07$  & $0.041$  \\\hline
 $\Delta(\si{\electronvolt})$ & $0.28$ & $0.34$ &  &  \\\hline
\end{tabular}
 \caption{Experimental measurements of the surface state gap, induced by coupling between opposite surfaces. First row is from Ref.~\onlinecite{experiment_gap}, second row is from Ref.~\onlinecite{experimental_gap_osc}. In the second row the gap increases from 2 to 3 QL, signifying an oscillatory dependence of the gap on the thickness.}
\label{table:experimental_gap}
\end{table}

\FloatBarrier

\section{Conclusion}\label{conclusion}
We have derived the most general Hamiltonian for the  $\text{Bi}_2\text{Se}_3$-class of materials including terms to third order in the wave vector, first order in electric and magnetic fields, first order in strain and first order in both strain and wave vector. We show that this model provides a description of a range of different effects of strain on the electronic structure of these materials. Specifically we have analytically derived the spectrum of the surface states for a semi-infinite topological insulator, showing qualitatively the effects of strain on both the spectrum and the spin structure. The terms first order in both wave vector and strain break the full rotational symmetry close to $k_{||}=0$, leading to an anisotropy in the Dirac cone otherwise absent for the simple (111) surface termination. The spin structure is altered as well, and for some strain configurations  a spin component perpendicular to the surface arises. In an analysis of the finite size effect we show that increasing the bulk band gap by virtue of strain, decreases the induced gap in the surface states. The possibility of tuning the SS gap gives new possibilities for experimental investigation of the thickness dependence of the SS gap and control of transport and optical properties.

\begin{acknowledgments}
MRJ and MW gratefully acknowledge financial support from the Danish Council of Independent Research (Natural Sciences) grant no.: DFF-4181-00182. AML gratefully acknowledges financial support from the Carlsberg Foundation.  The Center for Quantum Devices is funded by the Danish National Research Foundation.
\end{acknowledgments}

\appendix

\section{Effective masses}
\label{eff_mass}
Here we give the expressions for the bulk effective masses including both contributions from strain and electric fields. Upper (lower) signs refer to the conduction (valence) band.
\begin{widetext}
\begin{align}
\frac{\hbar^2}{m_x}   &= 2 D_2 \pm \frac{2}{\Delta_\Gamma}\bigg(-2 (M + M_1\epsilon_{zz} + M_2\epsilon_{||})B_2 + ( A_2 + A_{21}\epsilon_{zz} + A_{22}\epsilon_{||}  +  Y_3 \epsilon_{zy} + Y_4(\epsilon_{xx} -\epsilon_{yy}))^2 \nonumber\\
  &+ (Y_3\epsilon_{zx} + 2 Y_4\epsilon_{xy})^2 + (X_1 \epsilon_{zx} + 2 X_2\epsilon_{xy})^2 + ( -X_3 \epsilon_{zy} + X_4 (\epsilon_{xx}-\epsilon_{yy}))^2 \nonumber\\
 &-\frac{4}{\Delta_\Gamma^2} \Big( 4E_z^2 W_z^2 \big((A_2 + A_{21}\epsilon_{zz} + A_{22}\epsilon_{||}  +  Y_3 \epsilon_{zy} + Y_4(\epsilon_{xx} -\epsilon_{yy}))^2 + (Y_3\epsilon_{zx} + 2 Y_4\epsilon_{xy} )^2\big) \nonumber\\
 &+ 4 W_{||}^2 E_{||}^2 \big((X_1 \epsilon_{zx} + 2 X_2\epsilon_{xy})^2 + ( -X_3 \epsilon_{zy} + X_4 (\epsilon_{xx}-\epsilon_{yy}))^2\big) \nonumber\\
 &+ \big( W_{||} E_x (A_2 + A_{21}\epsilon_{zz} + A_{22}\epsilon_{||}  +  Y_3 \epsilon_{zy} + Y_4(\epsilon_{xx} -\epsilon_{yy})) - W_{||} E_y (Y_3\epsilon_{zx} + 2 Y_4\epsilon_{xy}) \nonumber\\
  &+ W_z E_z (X_1 \epsilon_{zx} + 2 X_2\epsilon_{xy}) \big)^2 \nonumber\\
  &+ \big( W_{||} E_x (Y_3\epsilon_{zx} + 2 Y_4\epsilon_{xy}) + W_{||} E_y (A_2 + A_{21}\epsilon_{zz} + A_{22}\epsilon_{||}  +  Y_3 \epsilon_{zy} + Y_4(\epsilon_{xx} -\epsilon_{yy})) \nonumber\\
   &+ W_z E_z ( -X_3 \epsilon_{zy} + X_4 (\epsilon_{xx}-\epsilon_{yy})) \big)^2 \nonumber\\
   &- 2\big( W_{||} E_x (A_2 + A_{21}\epsilon_{zz} + A_{22}\epsilon_{||}  +  Y_3 \epsilon_{zy} + Y_4(\epsilon_{xx} -\epsilon_{yy})) - W_{||} E_y (Y_3\epsilon_{zx} + 2 Y_4\epsilon_{xy}) \nonumber\\
    &- W_z E_z ( X_1 \epsilon_{zx} + 2 X_2\epsilon_{xy}) \big)^2 \nonumber\\
  &+ 2\big( W_{||} E_x (Y_3\epsilon_{zx} + 2 Y_4\epsilon_{xy}) + W_{||} E_y (A_2 + A_{21}\epsilon_{zz} + A_{22}\epsilon_{||}  +  Y_3 \epsilon_{zy} + Y_4(\epsilon_{xx} -\epsilon_{yy})) \nonumber\\
  &- W_z E_z (-X_3 \epsilon_{zy} + X_4 (\epsilon_{xx}-\epsilon_{yy})) \big)^2
  \Big) \bigg),
\end{align}

\begin{align}
\frac{\hbar^2}{m_y}
 &= 2 D_2 \pm \frac{2}{\Delta_\Gamma}\bigg(-2 (M + M_1\epsilon_{zz} + M_2\epsilon_{||}) B_2 + ( Y_3\epsilon_{zx} + 2 Y_4\epsilon_{xy})^2 \nonumber\\
 &+(A_2 + A_{21}\epsilon_{zz} + A_{22}\epsilon_{||}  -  Y_3 \epsilon_{zy} -Y_4(\epsilon_{xx} -\epsilon_{yy}))^2 + ( X_1\epsilon_{zy} +  X_2 (\epsilon_{xx}-\epsilon_{yy}))^2 + (X_3 \epsilon_{zx} - 2 X_4 \epsilon_{xy})^2 \nonumber\\
 &-\frac{4}{\Delta_\Gamma^2} \Big( 4 E_z^2 W_z^2 \big(( Y_3\epsilon_{zx} + 2 Y_4\epsilon_{xy})^2 + (A_2 + A_{21}\epsilon_{zz} + A_{22}\epsilon_{||}  -  Y_3 \epsilon_{zy} -Y_4(\epsilon_{xx} -\epsilon_{yy}))^2 \big) \nonumber\\
 &+ 4 W_{||}^2 E_{||}^2 \big( ( X_1\epsilon_{zy} +  X_2 (\epsilon_{xx}-\epsilon_{yy}))^2 + (X_3 \epsilon_{zx} - 2 X_4 \epsilon_{xy})^2 \big) \nonumber\\
 &+ \big( W_{||} E_x ( Y_3\epsilon_{zx} + 2 Y_4\epsilon_{xy}) - W_{||} E_y (A_2 + A_{21}\epsilon_{zz} + A_{22}\epsilon_{||}  -  Y_3 \epsilon_{zy} -Y_4(\epsilon_{xx} -\epsilon_{yy})) \nonumber\\
  &+ W_z E_z ( X_1\epsilon_{zy} +  X_2 (\epsilon_{xx}-\epsilon_{yy}))\big)^2 \nonumber\\
  &+ \big( W_{||} E_x (A_2 + A_{21}\epsilon_{zz} + A_{22}\epsilon_{||}  -  Y_3 \epsilon_{zy} -Y_4(\epsilon_{xx} -\epsilon_{yy})) + W_{||} E_y ( Y_3\epsilon_{zx} + 2 Y_4\epsilon_{xy}) \nonumber\\
   &+ W_z E_z (X_3 \epsilon_{zx} - 2 X_4 \epsilon_{xy})\big)^2 \nonumber\\
   &- 2\big( W_{||} E_x ( Y_3\epsilon_{zx} + 2 Y_4\epsilon_{xy}) - W_{||} E_y (A_2 + A_{21}\epsilon_{zz} + A_{22}\epsilon_{||}  -  Y_3 \epsilon_{zy} -Y_4(\epsilon_{xx} -\epsilon_{yy})) \nonumber\\
   &- W_z E_z ( X_1\epsilon_{zy} +  X_2 (\epsilon_{xx}-\epsilon_{yy}))\big)^2 \nonumber\\
  &+ 2\big( W_{||} E_x (A_2 + A_{21}\epsilon_{zz} + A_{22}\epsilon_{||}  -  Y_3 \epsilon_{zy} -Y_4(\epsilon_{xx} -\epsilon_{yy})) + W_{||} E_y ( Y_3\epsilon_{zx} + 2 Y_4\epsilon_{xy}) \nonumber\\
   &- W_z E_z (X_3 \epsilon_{zx} - 2 X_4 \epsilon_{xy})\big)^2
  \Big) \bigg),
\end{align}

\begin{align}
\frac{\hbar^2}{m_z}
&= 2 D_1 \pm \frac{2}{\Delta_\Gamma}\bigg(- 2  (M + M_1\epsilon_{zz} + M_2\epsilon_{||})B_1+ (Y_1\epsilon_{zx} + 2 Y_2 \epsilon_{xy})^2 + ( Y_1 \epsilon_{zy} + Y_2(\epsilon_{xx}-\epsilon_{yy}))^2 \nonumber\\
&+ (A_1 + A_{11}\epsilon_{zz} + A_{12}\epsilon_{||})^2 \nonumber\\
 &-\frac{4}{\Delta_\Gamma^2} \Big( 4 E_z^2 W_z^2 \big( (Y_1\epsilon_{zx} + 2 Y_2 \epsilon_{xy})^2 + ( Y_1 \epsilon_{zy} + Y_2(\epsilon_{xx}-\epsilon_{yy}))^2\big)
 + 4 W_{||}^2 E_{||}^2 ( A_1 + A_{11}\epsilon_{zz} + A_{12}\epsilon_{||}) \nonumber\\
 &+ \big( W_{||} E_x (Y_1\epsilon_{zx} + 2 Y_2 \epsilon_{xy}) - W_{||} E_y ( Y_1 \epsilon_{zy} + Y_2(\epsilon_{xx}-\epsilon_{yy})) + W_z E_z ( A_1 + A_{11}\epsilon_{zz} + A_{12}\epsilon_{||}) \big)^2 \nonumber\\
  &+ \big( W_{||} E_x ( Y_1 \epsilon_{zy} + Y_2(\epsilon_{xx}-\epsilon_{yy})) + W_{||} E_y (Y_1\epsilon_{zx} + 2 Y_2 \epsilon_{xy}) \big)^2 \nonumber\\
   &- 2\big( W_{||} E_x (Y_1\epsilon_{zx} + 2 Y_2 \epsilon_{xy}) - W_{||} E_y ( Y_1 \epsilon_{zy} + Y_2(\epsilon_{xx}-\epsilon_{yy})) - W_z E_z ( A_1 + A_{11}\epsilon_{zz} + A_{12}\epsilon_{||})\big)^2 \nonumber\\
  &+ 2\big( W_{||} E_x ( Y_1 \epsilon_{zy} + Y_2(\epsilon_{xx}-\epsilon_{yy})) + W_{||} E_y (Y_1\epsilon_{zx} + 2 Y_2 \epsilon_{xy}) \big)^2
  \Big) \bigg).
\end{align}
\end{widetext}

\bibliography{references}{}
\bibliographystyle{apsrev4-1}

\end{document}